\documentclass[preprint,12pt]{elsarticle}

\usepackage{amssymb}

\journal{Journal of Geometry and Physics}

\begin{document}

\begin{frontmatter}



\title{The $gl(1|1)$ Lie superbialgebras}


\author{A. Eghbali }
\ead{a.eghbali@azaruniv.edu}

\author{A. Rezaei-Aghdam\corref{fn2}}
\ead{rezaei-a@azaruniv.edu}

\cortext[fn2]{Corresponding author.}

\address{Department of Physics, Faculty of science,
Azarbaijan Shahid Madani University, 53714-161, Tabriz, Iran}

\begin{abstract}
By direct calculations of matrix form of super Jacobi and mixed
super Jacobi identities which are obtained from adjoint
representation, and using the  automorphism supergroup of the
$gl(1|1)$ Lie superalgebra, we determine and classify all
$gl(1|1)$ Lie superbialgebras. Then, by calculating their
classical r-matrices, the $gl(1|1)$ coboundary Lie superbialgebras
and their types (triangular, quasi-triangular or factorizable) are
determined, furthermore in this way super Poisson structures on
the $\bf GL(1|1)$ Lie supergroup are obtained. Also,  we classify
Drinfeld superdoubles based on the $gl(1|1)$ as a theorem.
Afterwards, as a physical application of the coboundary Lie
superbialgebras, we construct a new integrable system on the
homogeneous superspace $OSp(1|2)/U(1)$. Finally, we make use of
the Lyakhovsky and Mudrov formalism  in order to build up the
deformed $gl(1|1)$ Lie superalgebra related to all $gl(1|1)$
coboundary Lie superbialgebras. For one case, the quantization at
the supergroup level is also provided, including its quantum
R-matrix.
\end{abstract}


\end{frontmatter}


\section{Introduction}
\label{intro.}

From the mathematical point of view, the classification of Lie
bialgebras can be seen as the first step in the classification of
quantum groups. Many interesting examples of Lie bialgebras based
on complex semisimple Lie algebras have been given by Drinfeld
\cite{Drin}. A complete classification of Lie bialgebras with
reduction was given in \cite{Del}. However, a classification of
Lie bialgebras is out of reach, with similar reasons as for Lie
algebra classification. In the non-semisimple case, only a bunch
of low dimensional examples have been thoroughly studied
\cite{{JR},{RHR},{Gomez}}. On the other hand, from the physical
point of view, the theory of classical integrable systems
naturally relates to the geometry and representation theory of
Poisson-Lie groups and the corresponding Lie bialgebras and their
classical r-matrices (see for example \cite{Kosmann}). In the same
way, Lie superbialgebras \cite{N.A}, as the underlying symmetry
algebras, play an important role in the integrable structure of
$AdS/CFT$ correspondence \cite{Bs}. Similarly, one can consider
Poisson-Lie T-dual sigma models on Poisson-Lie supergroups
\cite{ER}. In this way, and by considering that there is a
universal quantization for Lie superbialgebras \cite{Geer}, one
can assign an important role to the classification of Lie
superbialgebras (especially low dimensional Lie superbialgebras)
from both, physical and mathematical point of view. Until now
there were distinguished and nonsystematic ways for obtaining low
dimensional Lie superbialgebras (see for example
\cite{{J.z},{J}}). In the previous paper, we presented a
systematic way for obtaining and classificating  low dimensional
Lie superbialgebras using the adjoint representation of Lie
superalgebras \cite{ER1}. Here, we apply this method to the
classification of the $gl(1|1)$ Lie superbialgebras. Note that the
classification of the quantum deformations of the $gl(1|1)$ Lie
superalgebra was obtained previously in Ref. \cite{Frappat}.
Indeed here, we present the classical version of that work in
another method. Furthermore, the physical models on the $\bf
GL(1|1)$ Lie supergroup such as WZW model and its relation  to
logarithmic CFT have been recently studied \cite{sh} (see, also
\cite{Creutzig}). So, for these reasons, obtaining of the
$gl(1|1)$ Lie superbialgebras is the first step in the
presentation of super Poisson-Lie symmetry \cite{ER} in such
models.

\smallskip
\goodbreak

The paper is organized as follows. In section two, we review  some
basic definitions and notations that are used throughout the
paper. Matrix form of super Jacobi  and mixed super Jacobi
identities \cite{ER1} of Lie superbialgebras is rewritten in that
section. A list of indecomposable four-dimensional Lie
superalgebras of the type $(2 , 2)$  \cite{B} is offered in
section three. Furthermore, in that section  we give the complete
list of decomposable Lie superalgebras of the type $(2,2)$. The
automorphism Lie supergroup of the $gl(1|1)$ Lie superalgebra is
also presented in section three. Then, using the method mentioned
in  \cite{ER1}, we classify all  $gl(1|1)$ Lie superbialgebras in
section four; the details of calculations are also explained. In
section five, by calculating the classical r-matrices of the
$gl(1|1)$ Lie superbialgebras, we determine their types,
furthermore in this way, we obtain all super Poisson structures on
the $\bf GL(1|1)$ Lie supergroup. Section six is devoted to the
classification of the Drinfeld superdoubles based on  the
$gl(1|1)$; the isomorphism matrices between all the  Manin
supertriples generated by $gl(1|1) $ have given in  Appendix . In
section seven, by using the classical r-matrices of the $gl(1|1)$
we construct a new integrable system on a supersymplectic
supermanifold, namely the homogeneous superspace $OSp(1|2)/U(1)$.
Finally, in section eight by using the Lyakhovsky and Mudrov
formalism \cite{Lyakhovsky}, we quantize $gl(1|1)$ Lie
superalgebra in related to its various coboundary Lie
superbialgebras, and at the end, for one case the quantization of
the supergroup including quantum R-matrix  is provided. Some
remarks are given in the conclusion section.


\section{Definitions and notations}

Here, we apply DeWitt notations for supervector spaces,
supermatrices, etc \cite{D}. In this way, for self-containing of
the paper, we recall some basic definitions and a proposition on
Lie superbialgebras \cite{{RHR},{N.A},{ER1}}.

{\bf Definition 1:} A  {\it Lie superalgebra} $\mathcal{G}$ is a
graded vector space $\mathcal{G}=\mathcal{G}_B \oplus
\mathcal{G}_F$ with gradings $grade(\mathcal{G}_B)=0$ and
$grade(\mathcal{G}_F)=1$  so that Lie bracket satisfies the super
antisymmetric and super Jacobi identities, i.e., in the graded
basis $\{X_i\}$ of $\mathcal{G}$ we have \footnote{ Note that the
bracket of one boson with one boson or one fermion is usual
commutator, but for one fermion with one fermion is
anticommutator. Furthermore, we identify grading of indices by the
same indices in the power of (-1), for example $grading(i)\equiv
i$; this is the notation that DeWitt applied in \cite{D}.
Meanwhile we work in the standard basis \cite{D}, i.e., in writing
the basis of Lie superalgebras, we consider bosonic generators
before fermionic ones. }
\begin{equation}
[X_i , X_j] = {f^k}_{ij} X_k,\label{1}
\end{equation}
and
\begin{equation}
(-1)^{i(j+k)}{f^m}_{jl}{f^l}_{ki} + {f^m}_{il}{f^l}_{jk} +
(-1)^{k(i+j)}{f^m}_{kl}{f^l}_{ij}=0,\label{2}
\end{equation}
so that
\begin{equation}
{f^k}_{ij}=-(-1)^{ij}{f^k}_{ji}.\label{3}
\end{equation}
Furthermore, we have
\begin{equation}
{f^k}_{ij}=0, \hspace{10mm} if \hspace{5mm} grade(i) +
grade(j)\neq grade(k)\hspace{3mm} (mod 2).\label{4}
\end{equation}
Note that using the adjoint representation
\begin{equation}
({{\cal X}}_i)_j^{\; \;k} = -{f^k}_{ij}, \hspace{10mm} ({\cal
Y}^i)_{ \;jk} = -{f^i}_{jk},\label{5}
\end{equation}
the super Jacobi identities can be rewritten in the following
matrix form \cite{ER1}:
\begin{equation}
({{\cal X}}_i)_j^{\; \;k}{{\cal X}}_k - {{\cal X}}_j {{\cal X}}_i
+ (-1)^{ij}{{\cal X}}_i {{\cal X}}_j = 0.\label{6}
\end{equation}
Let $\mathcal{G}$ be a finite-dimensional Lie superalgebra and
$\mathcal{G}^\ast$ be its dual superspace with respect to a
non-degenerate canonical pairing $( . ~,~. )$ on $\mathcal{G}^\ast
\oplus \mathcal{G}$.

{\bf Definition 2:} A {\it Lie superbialgebra} structure on a Lie
superalgebra $\mathcal{G}$ is a linear map $\delta :
\mathcal{G}\longrightarrow \mathcal{G} \otimes \mathcal{G}$ (the
{\em super cocommutator}) so that

1)~~$\delta$ is a super one-cocycle, i.e.,
\begin{equation}
\delta([X,Y])=(ad_X\otimes I+I\otimes
ad_X)\delta(Y)-(-1)^{|X||Y|}(ad_Y\otimes I+I\otimes
ad_Y)\delta(X),\label{7}
\end{equation}
~~~~~~~~where $X,Y\in {\bf \mathcal{G}}$ and {\footnotesize
$|X|(|Y|)$} indicates the grading of $X(Y)$.

2) The dual map ${^t}{\delta}:{\bf \mathcal{G}}^\ast\otimes {\bf
\mathcal{G}}^\ast \to {\bf \mathcal{G}}^\ast$ is a Lie
superbracket on ${\bf \mathcal{G}}^\ast$, i.e.,
\begin{equation}
(\xi\otimes\eta , \delta(X)) = ({^t}{\delta}(\xi\otimes\eta) , X)
= ([\xi,\eta]_\ast , X),    \qquad \forall X\in  {\bf \mathcal{G}}
;\;\, \xi,\eta\in{\bf \mathcal{G}}^\ast.\label{8}
\end{equation}
The Lie superbialgebra defined in this way will be denoted by
$({\bf \mathcal{G}},{\bf \mathcal{G}}^\ast)$ or $({\bf
\mathcal{G}},\delta)$ \cite {{N.A},{ER1}}.

{\bf Definition 3:} A Lie superbialgebra is  {\it coboundary} if
the super cocommutator is a one-coboundary, i.e., if there exists
an element $r\in{\bf \mathcal{G}}\otimes{\bf \mathcal{G}}$ such
that
\begin{equation}
\delta(X) = (I\otimes ad_X + ad_X\otimes I)r,        \qquad
\forall X\in {\bf \mathcal{G}}.\label{9}
\end{equation}

{\bf Proposition 1:} Two coboundary Lie superbialgebras $({\bf
\mathcal{G}},{\bf \mathcal{G}}^\ast)$ and $({\bf
\mathcal{G}}^\prime,{{\bf \mathcal{G}}^\ast}^\prime)$ defined by
$r\in \bf \mathcal{G} \otimes \bf \mathcal{G}$ and $r^\prime \in
{\bf \mathcal{G}}^\prime \otimes {\bf \mathcal{G}}^\prime$ are
{\em isomorphic} if and only if there is an isomorphism of Lie
superalgebras $\alpha: \bf \mathcal{G} \longrightarrow {\bf
\mathcal{G}}^\prime$ such that $(\alpha\otimes\alpha)r - r^\prime$
is ${\bf \mathcal{G}}^\prime$ invariant \cite{RHR}, i.e.,
\begin{equation}
(I\otimes ad_X + ad_X\otimes I)((\alpha\otimes\alpha)r -
r^\prime)=0, \qquad \forall X\in {\bf
\mathcal{G}}^\prime.\label{10}
\end{equation}

{\bf Definition 4:} Coboundary Lie superbialgebras can be of two different types:\\

$(a)$ if we denote $r=r^{ij}X_i \otimes X_j$ and  $r$  be a  super
skew-symmetric solution ($r^{ij} = -(-1)^{ij} r^{ji}$) of the
graded classical Yang-Baxter equation (GCYBE)
\begin{equation}
[[r , r]] = 0,\label{11}
\end{equation}
then the coboundary Lie superbialgebra is said to be {\it
triangular}; where the  Schouten superbracket is defined by
\begin{equation}
[[r , r]] = [r_{12}, r_{13}] + [r_{12} , r_{23}] + [r_{13} ,
r_{23}],\label{12}
\end{equation}
such that $r_{12}= r^{ij}X_i \otimes X_j \otimes 1$, $r_{13}=
r^{ij}X_i \otimes 1 \otimes X_j$ and $r_{23}= r^{ij}1 \otimes X_i
\otimes X_j$. A solution of the GCYBE is often called a {\it
classical r-matrix}. With regard to the fact $r$ has even
Grassmann parity and Grassmann parity of $r^{ij}$ comes from
indices, i.e., $r^{ij}=0$ if $i+j=1$, one can show that
$$
[r_{12},r_{13}]=(-1)^{i(k+l)+jl}\;r^{ij}r^{kl}\;[X_i,X_k] \otimes
X_j \otimes X_l,\label{11}
$$
$$
~[r_{12},r_{23}]=(-1)^{(i+j)(k+l)}\;r^{ij}r^{kl}\;X_i \otimes
[X_j,X_k] \otimes X_l,\label{12}
$$
$$
[r_{13},r_{23}]=(-1)^{i(k+l)+jl}\;r^{ij}r^{kl}\;X_i \otimes X_k
\otimes [X_j,X_l],\label{13}
$$

$(b)$ if $r$ is a solution of GCYBE such that $r_{12} + r_{21}$ is
a ${\bf \mathcal{G}}$ invariant element of ${\bf
\mathcal{G}}\otimes{\bf \mathcal{G}}$, then the coboundary Lie
superbialgebra is said to be {\it quasi-triangular}. Moreover, if
the super symmetric part of $r$ is invertible, then $r$ is called
{\it factorizable}.

Sometimes condition $(b)$ can be replaced with the following:

$(b^\prime)$ if $r$ is a super skew-symmetric solution of the
modified GCYBE
\begin{equation}
[[r , r]] = \varpi,            \qquad \varpi\in {\wedge}^3 {\bf
\mathcal{G}},\label{13}
\end{equation}
then the coboundary Lie superbialgebra is said to be
quasi-triangular \cite{N.A}.\\
Note that if ${\bf \mathcal{G}}$ is a Lie superbialgebra then
${\bf \mathcal{G}}^\ast $ is also a Lie superbialgebra \cite{N.A},
but this is not always true for the coboundary property.

{\bf Definition 5:} A {\it Manin supertriple}  is a triple of Lie
superalgebras $(\cal{D} , {\bf \mathcal{G}} , {\bf
\tilde{\mathcal{G}}})$ together with a nondegenerate ad-invariant
supersymmetric bilinear form $<.~ , ~.
>$ on $\cal{D}$, such that\hspace{2mm}

(1)~~${\bf \mathcal{G}}$ and ${\bf \tilde{\mathcal{G}}}$ are Lie
subsuperalgebras of $\cal{D}$,\hspace{2mm}

(2)~~$\cal{D} = {\bf \mathcal{G}}\oplus{\bf \tilde{\mathcal{G}}}$
as a supervector space,\hspace{2mm}

(3)~~${\bf \mathcal{G}}$ and ${\bf \tilde{\mathcal{G}}}$ are
isotropic with respect to $< .~, ~. >$, i.e.,
$$
<X_i , X_j> = <\tilde{X}^i , \tilde{X}^j> = 0,
$$
\begin{equation}
{\delta_i}^j=\;<X_i , \tilde{X}^j>\; = (-1)^{ij}<\tilde{X}^j,
X_i>=(-1)^{ij}{\delta^j}_i,\label{17}
\end{equation}
where $\{X_i\}$ and $\{\tilde{X}^i\}$ are  basis of Lie
superalgebras ${\bf \mathcal{G}}$ and ${\bf \tilde{\mathcal{G}}}$,
respectively \cite {{N.A},{ER1}}.

Note that in the second equation of the (\ref{17}), ${\delta^j}_i$
is the ordinary delta function. There is a one-to-one
correspondence between Lie superbialgebra $({\bf \mathcal{G}},{\bf
\mathcal{G}}^\ast)$ and Manin supertriple $(\cal{D} , {\bf
\mathcal{G}} , {\bf \tilde{\mathcal{G}}})$ with ${\bf
\tilde{\mathcal{G}}}= {\bf \mathcal{G}}^\ast$. If we choose the
structure constants of Lie superalgebras ${\bf \mathcal{G}}$ and
${\bf \tilde{\mathcal{G}}}$ as
\begin{equation}
[X_i , X_j] = {f^k}_{ij} X_k,\hspace{20mm} [\tilde{X}^i ,\tilde{
X}^j] ={{\tilde{f}}^{ij}}_{\; \; \: k} {\tilde{X}^k},\label{18}
\end{equation}
then ad-invariance of the bilinear form $< .~ , ~. >$ on $\cal{D}
= {\bf \mathcal{G}}\oplus{\bf \tilde{\mathcal{G}}}$ implies that
\cite {ER1}
\begin{equation}
[X_i , \tilde{X}^j] =(-1)^j{\tilde{f}^{jk}}_{\; \; \; \:i} X_k
+(-1)^i {f^j}_{ki} \tilde{X}^k.\label{19}
\end{equation}
Clearly, using Eqs. (\ref{8}), (\ref{17}) and (\ref{19}) we have
\begin{equation}
\delta(X_i) = (-1)^{jk}{\tilde{f}^{jk}}_{\; \; \; \:i} X_j \otimes
X_k.\label{20}
\end{equation}
We note that the appearance of $(-1)^{jk}$ in this relation is due
to the definition of natural inner product between ${\bf
\mathcal{G}}\otimes {\bf \mathcal{G}}$ and  ${\bf
\mathcal{G}}^\ast \otimes {\bf \mathcal{G}}^\ast $ as $
(\tilde{X}^i \otimes \tilde{ X}^j ,X_k \otimes
X_l)=(-1)^{jk}{\delta^i}_k {\delta^j}_l$.\\
As a result, if we apply this relation in the super one-cocycle
condition (\ref{7}), the super Jacobi identities (\ref{2}) for the
dual Lie superalgebra and the following mixed super Jacobi
identities are obtained:
\begin{equation}
{f^m}_{jk}{\tilde{f}^{il}}_{\; \; \; \; m}=
{f^i}_{mk}{\tilde{f}^{ml}}_{\; \; \; \; \; j} +
{f^l}_{jm}{\tilde{f}^{im}}_{\; \; \; \; \; k}+ (-1)^{jl}
{f^i}_{jm}{\tilde{f}^{ml}}_{\; \; \; \; \; k}+ (-1)^{ik}
{f^l}_{mk}{\tilde{f}^{im}}_{\; \; \; \; \; j}.\label{21}
\end{equation}
This relation can also be obtained from super Jacobi identity of
$\cal{D}$. As mentioned in Ref. \cite{ER1}, using the adjoint
representation, the matrix form of mixed super Jacobi identities
has the following form:
\begin{equation}
{({\tilde{\cal X}}^i)}^j_{\; \;l}\;{\cal Y}^l =-(-1)^{k}
({\tilde{\cal X}}^{st})^{j}\; {\cal Y}^i + {\cal Y}^j{\tilde{\cal
X}}^i - (-1)^{ij}{\cal Y}^i {\tilde{\cal X}}^j + (-1)^{k+ij}
({\tilde{\cal X}}^{st})^{i} \;{\cal Y}^j,\label{22}
\end{equation}
where index $k$ represents the column of matrix ${\tilde{\cal
X}}^{st}$ and superscript $st$ stands for supertranspose.

\section{Lie superalgebras of the type $(2 , 2)$}

In this section, we use the classification of four-dimensional Lie
superalgebras of the type $(2 , 2)$  listed in \cite{B}. In that
classification, Lie superalgebras are divided into two types:
trivial and nontrivial Lie superalgebras for which the commutation
of fermion-fermion is zero or nonzero, respectively. (As we use
DeWitt notation here, the structure constant $C^B_{FF}$ must be
pure imaginary in the standard basis.)  The results have been
presented in tables I and II. Note that in \cite{B} only the
indecomposable Lie superalgebras are classified. Here we also
consider decomposable Lie superalgebras of the type $(2 , 2)$ as
presented in table III. As the tables  indicate, the Lie
superalgebras have two, $ \{X_1,X_2\}$ bosonic and  two,
$\{X_3,X_4\}$ fermionic generators. The labeling of the trivial
Lie superalgebras is such that the letters A, B, C and D with
integral superscripts i and real subscripts p and q, denote the
equivalence classes of Lie superalgebras of dimension d, where d=
1 for A, d=2 for B, d=3 for C and  d=4 for D. The superscript i
and real subscripts  p and q denote the number of non isomorphic
Lie superalgebras and the Lie superalgebras parameters,
respectively.  For the nontrivial Lie superalgebras, we add  the
bracketed symbol to the corresponding trivial Lie superalgebra,
where necessary, an integral superscript and a real subscript.

Note that, as mentioned in Ref. \cite{ER4} the four-dimensional
Drinfel'd superdoubles are equal or isomorphic to some of the Lie
superalgebras of table II in the following way
$$
(C^3+A)\; \equiv \;((A_{1,1}+A), I_{(1 , 1)}),
$$
$$
(C^2_{-1}+A)\; \cong \;(B, I_{(1 , 1)})\; \cong
\;(B,(A_{1,1}+A))\; \cong \; (B,(A_{1,1}+A).i).
$$
So four-dimensional Drinfel'd superdoubles  have no new results
for table II. As mentioned above, we give all possible
decomposable Lie superalgebras of the type $(2 , 2)$ in  table
III. We note that the Lie superalgebras of $(2A_{1,1}+2A)^1$ and
$(C^1_{\frac{1}{2}} + A)$ are decomposable Lie superalgebras such
that in Backhouse's classification \cite{B} have been given in the
list of indecomposable Lie superalgebras.

\vspace{4mm}

\hspace{-0.5cm}{\footnotesize  TABLE I.} {\small  Trivial
indecomposable Lie superalgebras of
the type $(2 , 2)$.}\\
    \begin{tabular}{l l l   p{15mm} }
    \hline\hline
{\scriptsize ${ \mathcal{G}}$ }& {\scriptsize Non-zero
commutation relations}&{\scriptsize Comments}  \smallskip\\
\hline
\smallskip

{\scriptsize $D^5$}& {\scriptsize $[X_1,X_3]=X_3, \;\;[X_1,X_4]=X_4,\;\;[X_2,X_4]=X_3$} \\

{\scriptsize $D^6$}&{\scriptsize
$[X_1,X_3]=X_3, \;\;[X_1,X_4]=X_4,\;\;[X_2,X_3]=-X_4,\;\;[X_2,X_4]=X_3 $}\\

{\scriptsize $D^1_{pq}$}&{\scriptsize
$[X_1,X_2]=X_2, \;\;[X_1,X_3]=pX_3,\;\;[X_1,X_4]=qX_4 $}& {\scriptsize $pq \neq 0,\;\; p\geq q$}\\

{\scriptsize $D^8_{p}$}&{\scriptsize
$[X_1,X_2]=X_2, \;\;[X_1,X_3]=pX_3,\;\;[X_1,X_4]=X_3+pX_4 $}& {\scriptsize $p \neq 0$}\\

{\scriptsize $D^9_{pq}$}&{\scriptsize
$[X_1,X_2]=X_2, \;\;[X_1,X_3]=pX_3-qX_4,\;\;[X_1,X_4]=qX_3+pX_4 $}& {\scriptsize $q > 0$}\\

{\scriptsize $D^{10}_{p}$}&{\scriptsize
$[X_1,X_2]=X_2, \;\;[X_1,X_3]=(p+1)X_3,\;\;[X_1,X_4]=pX_4,\;\;[X_2,X_4]=X_3 $} \smallskip\\

\hline\hline
\end{tabular}

\vspace{3mm}

\subsection{The $gl(1|1)$ Lie superalgebra and its automorphism supergroup}

Traditional notation for the $(C^2_{-1}+A)$ Lie superalgebra is
the $gl(1|1)$. The $gl(1|1)$ Lie superalgebra is spanned by the
set of generators $\{ X_1, X_2, X_3, X_4 \}$ with grading;
$grade(X_1)=grade(X_2)=0$ and $grade(X_3)=grade(X_4)=1$, which in
the standard basis, fulfill the following (anti) commutation
relations
\begin{equation}
[X_1 , X_3] = X_3,\;\;\;[X_1 , X_4] = -X_4,\;\;\{X_3 , X_4
\}=iX_2.\label{23}
\end{equation}

 \hspace{-0.5cm}{\footnotesize TABLE II.} {\small
{ Non-trivial indecomposable Lie superalgebras of the type $(2 , 2).$}}\\
    \begin{tabular}{l l l  l p{15mm} }
    \hline\hline
{\scriptsize $ \mathcal{G}$ }& {\scriptsize Non-zero (anti)
commutation relations}&{\scriptsize Comments}  \smallskip\\
\hline
\smallskip

\vspace{-2mm}

{\scriptsize ${(D^7_{\frac{1}{2}\;\frac{1}{2}})}^1$}& {\scriptsize
$[X_1,X_2]=X_2, \;\;[X_1,X_3]=\frac{1}{2}
X_3,\;\;[X_1,X_4]=\frac{1}{2}
X_4,\;\; \{X_3,X_3\}=iX_2,$} \\

& {\scriptsize
$\{X_4,X_4\}=iX_2$} \\

\vspace{-2mm}

{\scriptsize ${(D^7_{\frac{1}{2}\;\frac{1}{2}})}^2$}& {\scriptsize
$[X_1,X_2]=X_2, \;\;[X_1,X_3]=\frac{1}{2}
X_3,\;\;[X_1,X_4]=\frac{1}{2}
X_4,\;\; \{X_3,X_3\}=iX_2,$} \\

& {\scriptsize
$ \{X_4,X_4\}=-iX_2$} \\

{\scriptsize ${(D^7_{\frac{1}{2}\;\frac{1}{2}})}^3$}& {\scriptsize
$[X_1,X_2]=X_2, \;\;[X_1,X_3]=\frac{1}{2}
X_3,\;\;[X_1,X_4]=\frac{1}{2}
X_4,\;\; \{X_3,X_3\}=iX_2 $} \\

{\scriptsize $(D^7_{1-p\;p})$ }& {\scriptsize $[X_1,X_2]=X_2,
\;\;[X_1,X_3]=p X_3,\;\;[X_1,X_4]=(1-p)
X_4,\;\{X_3,X_4\}=iX_2 $}& {\scriptsize $p \leq \frac{1}{2}$} \\

{\scriptsize $(D^8_{\frac{1}{2}})$ }& {\scriptsize $[X_1,X_2]=X_2,
\;\;[X_1,X_3]=\frac{1}{2} X_3,\;\;[X_1,X_4]=X_3+\frac{1}{2}
X_4,\; \{X_4,X_4\}=iX_2 $} \\

\vspace{-1mm}

{\scriptsize $(D^9_{\frac{1}{2}\;p})$ }& {\scriptsize
$[X_1,X_2]=X_2, \;\;[X_1,X_3]=\frac{1}{2} X_3-p
X_4,\;\;[X_1,X_4]=p X_3+\frac{1}{2} X_4,$}&
 {\scriptsize $p > 0$}  \\

& {\scriptsize $\{X_3,X_3\}=iX_2,\;\; \{X_4,X_4\}=iX_2$}&
  \\

\vspace{-1mm}

{\scriptsize $(D^{10}_{0})^1$ }& {\scriptsize $[X_1,X_2]=X_2,
\;\;[X_1,X_3]= X_3, \;\;[X_2,X_4]= X_3, \;\; \{X_4,X_4\}=iX_1,$}\\

&{\scriptsize $\{X_3,X_4\}=-i\frac{1}{2}X_2$}\\

\vspace{-1mm}

{\scriptsize $(D^{10}_{0})^2$ }& {\scriptsize $[X_1,X_2]=X_2,
\;\;[X_1,X_3]= X_3, \;\;[X_2,X_4]= X_3, \;\; \{X_4,X_4\}=-iX_1,$}\\

&{\scriptsize $\{X_3,X_4\}=i\frac{1}{2}X_2$}\\

{\scriptsize $(2A_{1,1}+2A)^2$ }& {\scriptsize $\{X_3,X_3\}=iX_1,
\;\; \{X_4,X_4\}=iX_2, \;\; \{X_3,X_4\}=iX_1 $}&
 {\scriptsize Nilpotent}\\

{\scriptsize ${(2A_{1,1}+2A)}^3_p$ }& {\scriptsize
$\{X_3,X_3\}=iX_1, \;\; \{X_4,X_4\}=iX_2, \;\;
\{X_3,X_4\}=ip(X_1+X_2)  $}
& {\scriptsize $p > 0\;$ Nilpotent}\\

{\scriptsize ${(2A_{1,1}+2A)}^4_p$ }& {\scriptsize
$\{X_3,X_3\}=iX_1, \;\; \{X_4,X_4\}=iX_2, \;\;
\{X_3,X_4\}=ip(X_1-X_2)  $}
& {\scriptsize $p > 0\;$ Nilpotent}\\

{\scriptsize $(C^1_1+A)$ }& {\scriptsize $[X_1,X_2]=X_2,
\;\;[X_1,X_3]=X_3,
\;\;\{X_3,X_4\}=iX_2 $} \\

{\scriptsize $(C^2_{-1}+A)$ }& {\scriptsize $[X_1,X_3]=X_3,
\;\;[X_1,X_4]=-X_4,
\;\;\{X_3,X_4\}=iX_2 $}& {\scriptsize Jordan-Winger }\\

& & {\scriptsize quantization }\\

{\scriptsize $(C^3+A)$ }& {\scriptsize $[X_1,X_4]=X_3,
\;\;\{X_4,X_4\}=iX_2 $}&{\scriptsize Nilpotent}\\

{\scriptsize $(C^5_0+A)$ }& {\scriptsize $[X_1,X_3]=-X_4,
\;\;[X_1,X_4]=X_3,
\;\;\{X_3,X_3\}=iX_2\;\;\{X_4,X_4\}=iX_2 $}\smallskip\\

\hline\hline
\end{tabular}
 {\footnotesize   The Lie superalgebra $A$ is  one
dimensional Abelian Lie superalgebra with one fermionic generator
where Lie superalgebra $A_{1,1}$ is its bosonization. }

\smallskip

To obtain the dual Lie superalgebras for the $gl(1|1)$, we need
automorphism supergroup of this Lie superalgebra. In order to
calculate the automorphism supergroup of the $gl(1|1)$ Lie
superalgebra, we use the following transformation \cite{ER1}:
\begin{equation}
{X'_i}=(-1)^j\;A_i^{\;\;j} X_j,\hspace{20mm} [X'_i ,X'_j] =f^k_{\;
\; ij} X'_k,\label{24}
\end{equation}
thus, we have the following matrix equation for the elements of
automorphism supergroup \cite{ER1}
\begin{equation}
(-1)^{ij+mk} A {\cal Y}^k A^{st} = {\cal Y}^e
A_e^{\;\;k},\label{25}
\end{equation}
where the indices $i$ and $j$ correspond to the row and column of
matrix ${\cal Y}^k$, respectively, and $m$ denotes the column of
matrix $A^{st}$  in the left hand side. As the Lie superalgebra is
generated by just two bosons $X_1$ and $X_2$, every automorphism
is determined by the following transformation of the bosons:
\begin{equation}
{X^{'}}_1=X_1+aX_2,~~~~{X^{'}}_2=bcX_2.
\end{equation}
For two fermions $X_3$ and $X_4$, we find that under the above
transformation
\begin{equation}
{X^{'}}_3=-bX_3,~~~~{X^{'}}_4=-cX_4,
\end{equation}
where $a \in R$ and $b, c \in R-\{0\}$. Finally, using Eq.
(\ref{25}) one can derive the matrix form of automorphism
supergroup of the $gl(1|1)$ Lie superalgebra as follows:
\begin{equation}
A=\left( \begin{tabular}{cccc}
               1&  a & 0 &0\\
                 0&  bc & 0 &0\\
                0&  0 & b &0\\
                  0&  0 & 0 &c\\
                \end{tabular} \right).
\end{equation}

\section{The $gl(1|1)$ Lie superbialgebras}

To obtain the $gl(1|1)$ Lie superbialgebras, we must first solve
Eq. (\ref{6}) for dual Lie superalgebras and Eq. (\ref{22}),
then, by using automorphism  supergroup of the $gl(1|1)$ Lie
superalgebra and the method mentioned in \cite{ER1} we classify
the $gl(1|1)$ Lie superbialgebras. The solutions of super Jacobi
(Eq. (\ref{6}) for dual Lie superalgebras) and mixed super Jacobi
(\ref{22}) identities for dual  Lie superalgebras of the
$gl(1|1)$ have the following four forms:
\begin{itemize}

\item $Case\; A:$

\vspace{-2mm}
$$
{{\tilde{f}}^{33}}_{\; \; \: 1}= {{\tilde{f}}^{33}}_{\; \; \:
1},\;\;\;\;{{\tilde{f}}^{44}}_{\; \; \: 1}={{\tilde{f}}^{44}}_{\;
\; \: 1},\;\;\;\;{{\tilde{f}}^{23}}_{\; \; \: 4}=\frac{-i}{2}
{{\tilde{f}}^{33}}_{\; \; \: 1},\;\;\;\;{{\tilde{f}}^{24}}_{\; \;
\: 3}=\frac{i}{2} {{\tilde{f}}^{44}}_{\; \; \: 1}.
$$
\vspace{-1mm} In this case, the general forms of the (anti)
commutation relations for $\tilde {\mathcal{G}}$ are as follows:
\vspace{-1mm}
$$
[{\tilde X}^2 , {\tilde X}^3] = \frac{\alpha}{2} {\tilde X}^4,~~
[{\tilde X}^2 , {\tilde X}^4] = \frac{-\beta}{2} {\tilde
X}^3,~~\{{\tilde X}^3 , {\tilde X}^3\} = i \alpha {\tilde X}^1,
~~\{{\tilde X}^4 , {\tilde X}^4\} = i \beta {\tilde X}^1.
$$

\vspace{3mm}

\hspace{-0.5cm}{\footnotesize TABLE III.} {\small
{ Decomposable Lie superalgebras of the type $(2 , 2).$}}\\
    \begin{tabular}{l l l l l p{15mm} }
    \hline\hline
 &{\footnotesize $ \mathcal{G}$ }& {\footnotesize Non-zero
(anti)commutation relations}&{\footnotesize Comments}  \smallskip\\
\hline
\smallskip

\vspace{0.5mm}

{\scriptsize Trivial}&{\footnotesize $I_{(2 , 2)}$}& {\scriptsize
All of the (anti)commutation relations are zero.} \\

&{\footnotesize $B \oplus B$}& {\footnotesize $[X_1,X_3]=X_3,
\;\;[X_2,X_4]=
X_4$} \\

&{\footnotesize $C^1_{p} \oplus A$}& {\footnotesize
$[X_1,X_2]=X_2, \;\;[X_1,X_3]=p X_3$} &{\footnotesize
$p \in \Re -\{0\}$}\\

&{\footnotesize $C^2_{p} \oplus A_{1,1}$}& {\footnotesize
$[X_1,X_3]=X_3, \;\;[X_1,X_4]=p X_4$} &{\scriptsize
$0< |p| \leq 1 $}\\

&{\footnotesize $L \oplus 2A$}& {\footnotesize
$[X_1,X_2]=X_2$}&{\scriptsize
$\equiv C^1_{p=0} \oplus A$} \\

&{\footnotesize $B \oplus A \oplus A_{1,1}$}& {\footnotesize
$[X_1,X_3]=X_3$}&{\scriptsize
$\equiv C^2_{p=0} \oplus A_{1,1}$} \\

&{\footnotesize $C^3 \oplus A_{1,1}$}& {\footnotesize
$[X_1,X_4]=X_3$}& {\scriptsize Nilpotent} \\

\vspace{0.5mm}

&{\footnotesize $C^4 \oplus A_{1,1}$}& {\footnotesize
$[X_1,X_3]=X_3,\;\;\; [X_1,X_4]=X_3+X_4$}& \\

\vspace{0.5mm}

&{\footnotesize $C^5_p \oplus A_{1,1}$}& {\scriptsize
$[X_1,X_3]=pX_3-X_4,\;[X_1,X_4]=X_3+pX_4$}&{\scriptsize
$p \geq 0 $} \\

\vspace{-1mm}

{\scriptsize Nontrivial}&{\footnotesize $(2A_{1,1}+2A)^0$ }& {\footnotesize $\{X_3,X_3\}=iX_1  $}&
{\scriptsize $\equiv(A_{1,1}+A)\oplus A \oplus A_{1,1}$,}\\

&& &
{\scriptsize  Nilpotent}\\

\vspace{-1mm}

&{\footnotesize $(2A_{1,1}+2A)^1$ }& {\footnotesize $\{X_3,X_3\}=iX_1,\; \;\; \{X_4,X_4\}=iX_2  $}&
{\scriptsize $\equiv(A_{1,1}+A)\oplus (A_{1,1}+A)$,}\\

&&&
{\scriptsize Nilpotent}\\

\vspace{0.5mm}

&{\scriptsize $(B \oplus (A_{1,1}+A))$ }& {\footnotesize $[X_1,X_3]=X_3,\; \;\; \{X_4,X_4\}=iX_2  $}\\

\vspace{0.5mm}

&{\scriptsize $((A_{1,1}+2A)^1 \oplus A_{1,1}) $ }& {\footnotesize
$\{X_3,X_3\}=iX_1,\; \;\; \{X_4,X_4\}=iX_1  $}&
 {\scriptsize Nilpotent}\\

\vspace{0.5mm}

&{\scriptsize $((A_{1,1}+2A)^2 \oplus A_{1,1}) $ }& {\footnotesize
$\{X_3,X_3\}=iX_1,\; \;\; \{X_4,X_4\}=-iX_1  $}&
 {\scriptsize Nilpotent}\\

\vspace{-1mm}

&{\footnotesize $(C^1_{\frac{1}{2}} + A)$ }& {\footnotesize
$[X_1,X_2]=X_2,\; \;[X_1,X_3]=\frac{1}{2}X_3,
$}&{\scriptsize $\equiv(C^1_{\frac{1}{2}}) \oplus A$ }\\

&& {\footnotesize $\{X_3,X_3\}=iX_2
$}&\smallskip\\
\hline\hline

\end{tabular}\\
{\footnotesize  L is a two-dimensional bosonic Lie algebra and
$(A_{1,1}+A)$ is   regarded as a subsuperalgebra of
$sl(1|1)\subset gl(1|1)$. }\\

\item $Case\; B:$

\vspace{-2mm}
$$
{{\tilde{f}}^{23}}_{\; \; \: 3}= {{\tilde{f}}^{23}}_{\; \; \:
3},\;\;\;\;{{\tilde{f}}^{24}}_{\; \; \: 4}={{\tilde{f}}^{24}}_{\;
\; \: 4},
$$
\vspace{-1mm} where the general forms of the (anti) commutation
relations for $\tilde {\mathcal{G}}$ are given by \vspace{1mm}
$$
[{\tilde X}^2 , {\tilde X}^3] = \alpha {\tilde X}^3,~~~ [{\tilde
X}^2 , {\tilde X}^4] = \beta {\tilde X}^4.
$$

\item $Case\; C:$

\vspace{-2mm}
$$
{{\tilde{f}}^{44}}_{\; \; \: 1}= {{\tilde{f}}^{44}}_{\; \; \:
1},\;\;\;\;{{\tilde{f}}^{24}}_{\; \; \: 3}=\frac{i}{2}
{{\tilde{f}}^{44}}_{\; \; \: 1},\;\;\;\;{{\tilde{f}}^{23}}_{\; \;
\: 3}={{\tilde{f}}^{23}}_{\; \; \: 3}.
$$
\vspace{-1mm} In this case, the general forms of the (anti)
commutation relations for $\tilde {\mathcal{G}}$ are as follows:
\vspace{-1mm}
$$
[{\tilde X}^2 , {\tilde X}^3] = \alpha {\tilde X}^3,~~~[{\tilde
X}^2 , {\tilde X}^4] = \frac{-\beta}{2} {\tilde X}^3,~~~\{{\tilde
X}^4 , {\tilde X}^4\} = i \beta {\tilde X}^1.
$$

\item $Case\; D:$

\vspace{-2mm}
$$
{{\tilde{f}}^{33}}_{\; \; \: 1}= {{\tilde{f}}^{33}}_{\; \; \:
1},\;\;\;\;{{\tilde{f}}^{23}}_{\; \; \: 4}=\frac{-i}{2}
{{\tilde{f}}^{33}}_{\; \; \: 1},\;\;\;\;{{\tilde{f}}^{24}}_{\; \;
\: 4}={{\tilde{f}}^{24}}_{\; \; \: 4},
$$
\vspace{-1mm} where the general forms of the (anti) commutation
relations for $\tilde {\mathcal{G}}$ are given by \vspace{-1mm}
$$
[{\tilde X}^2 , {\tilde X}^4] = \beta {\tilde X}^4,~~~[{\tilde
X}^2 , {\tilde X}^3] = \frac{\alpha}{2} {\tilde X}^4,~~~\{{\tilde
X}^3 , {\tilde X}^3\} = i \alpha {\tilde X}^1.
$$
In all cases, $\alpha$ and $\beta$ are  real constants.
\end{itemize}
In the following, we find isomorphism matrices $C$ between the
above solutions and Lie superalgebras of tables I, II and III. In
this way, we see that the solutions of super Jacobi and mixed
super Jacobi identities for dual  Lie superalgebras of the
$gl(1|1)$ are isomorphic to the $(C^3+A),$ $(C^2_{-1}+A)$ and
$(C^5_{0}+A)$ Lie superalgebras of table II and the $B \oplus A
\oplus A_{1,1}, C^2_{p} \oplus A_{1,1}$ and $(B \oplus
(A_{1,1}+A))$ Lie superalgebras of table III.

{$\bullet$} The  solution of $Case \;A$ is isomorphic to the
$(C^3+A),$ $(C^2_{-1}+A)$ and $(C^5_{0}+A)$ Lie superalgebras such
that isomorphism matrices $C$ between this solution  and the
$(C^3+A)$ Lie superalgebra are as follows:
$$
C_1=\left( \begin{tabular}{cccc}
              $ c_{11}$ & $c_{12}$& $ 0$ & $ 0$\\
                $ -i{{\tilde{f}}^{44}}_{\; \; \: 1}{c_{44}}^2$ & 0 & 0 & $ 0$\\
                 0  & 0 & $ \frac{i{{\tilde{f}}^{44}}_{\; \; \: 1}}{2}c_{12}c_{44}$ & $ 0$\\
                 0  & 0 & $ c_{43}$ & $ c_{44}$\\
                 \end{tabular}
                 \right),\;c_{12}, c_{44}\in\Re-\{0\};\;c_{11}, c_{43}\in\Re,
$$\\
where ${{\tilde{f}}^{33}}_{\; \; \: 1}={{\tilde{f}}^{23}}_{\; \;
\: 4}=0$, and
$$
C_2=\left( \begin{tabular}{cccc}
              $ c_{11}$ & $c_{12}$& $ 0$ & $ 0$\\
                $ -i{{\tilde{f}}^{33}}_{\; \; \: 1}{c_{43}}^2$ & 0 & 0 & $ 0$\\
                 0  & 0 & $ 0$ & $ -\frac{i{{\tilde{f}}^{33}}_{\; \; \: 1}}{2}c_{12}c_{43}$\\
                 0  & 0 & $ c_{43}$ & $ c_{44}$\\
                 \end{tabular}
                 \right),\;c_{12}, c_{44}\in\Re-\{0\};\;c_{11}, c_{43}\in\Re,
$$\\
where ${{\tilde{f}}^{44}}_{\; \; \: 1}={{\tilde{f}}^{24}}_{\; \; \: 3}=0$.\\
Furthermore, the isomorphism matrix related to the $(C^2_{-1}+A)$
Lie superalgebra is given by
$$
C_3=\left( \begin{tabular}{cccc}
              $ c_{11}$ & $c_{12}$& $ 0$ & $ 0$\\
                $ -2i{{\tilde{f}}^{44}}_{\; \; \: 1}c_{34}c_{44}$ & 0 & 0 & $ 0$\\
                 0  & 0 & $ \frac{i{{\tilde{f}}^{44}}_{\; \; \: 1}}{2}c_{12}c_{34}$ & $c_{34}$\\
                 0  & 0 & $-\frac{i{{\tilde{f}}^{44}}_{\; \; \: 1}}{2}c_{12}c_{44}$ & $ c_{44}$\\
                 \end{tabular}
                 \right),\;c_{12},c_{34},c_{44}\in\Re-\{0\};\;c_{11}\in\Re,
$$\\
where ${{\tilde{f}}^{33}}_{\; \; \:
1}=(\frac{4}{{c_{12}}^2})\frac{1}{{{\tilde{f}}^{44}}_{\; \; \;
1}}$ and  finally the isomorphism matrix $C$ between this case
and the $(C^5_{0}+A)$ Lie superalgebra is as follows:
$$
C_4=\left( \begin{tabular}{cccc}
              $ c_{11}$ & $c_{12}$& $ 0$ & $ 0$\\
                ${\scriptsize -i{{\tilde{f}}^{44}}_{\; \; \: 1}({c_{34}}^2+{c_{44}}^2)}$ & 0 & 0 & $ 0$\\
                 0  & 0 & $ \frac{i{{\tilde{f}}^{44}}_{\; \; \: 1}}{2}c_{12}c_{44}$ & $c_{34}$\\
                 0  & 0 & $-\frac{i{{\tilde{f}}^{44}}_{\; \; \: 1}}{2}c_{12}c_{34}$ & $ c_{44}$\\
                 \end{tabular}
                 \right),
$$\\
where ${{\tilde{f}}^{33}}_{\; \; \:
1}=-(\frac{4}{{c_{12}}^2})\frac{1}{{{\tilde{f}}^{44}}_{\; \; \;
1}}$ and $c_{12},c_{34},c_{44}\in\Re-\{0\};\;c_{11}\in\Re$.

{$\bullet$} The  solution of $Case \;B$ is isomorphic to the $B
\oplus A \oplus A_{1,1}$ and $C^2_{p}\oplus A_{1,1}$ Lie
superalgebras such that isomorphism matrices $C$ between this
solution  and the $B \oplus A \oplus A_{1,1}$ Lie superalgebra are
as follows:
$$
C_1=\left( \begin{tabular}{cccc}
              $ c_{11}$ & $\frac{1}{{{\tilde{f}}^{23}}_{\; \; \: 3}}$& $ 0$ & $ 0$\\
                $c_{21}$ & 0 & 0 & $ 0$\\
                 0  & 0 & $c_{33}$ & $ 0$\\
                 0  & 0 & $ 0$ & $ c_{44}$\\
                 \end{tabular}
                 \right),\;\;\;\;c_{21},c_{33},c_{44}\in\Re-\{0\};\;\;c_{11}\in\Re,
$$\\
where ${{\tilde{f}}^{24}}_{\; \; \: 4}=0$, and
$$
C_2=\left( \begin{tabular}{cccc}
              $ c_{11}$ & $\frac{1}{{{\tilde{f}}^{24}}_{\; \; \: 4}}$& $ 0$ & $ 0$\\
                $c_{21}$ & 0 & 0 & $ 0$\\
                 0  & 0 & $0$ & $c_{34}$\\
                 0  & 0 & $c_{43}$ & $0$\\
                 \end{tabular}
                 \right),\;\;\;\;c_{21},c_{34},c_{43}\in\Re-\{0\};\;\;c_{11}\in\Re,
$$\\
where ${{\tilde{f}}^{23}}_{\; \; \: 3}=0$.\\
The isomorphism matrices related to the $C^2_{p}\oplus A_{1,1}$
Lie superalgebras are given by
$$
C_3=\left( \begin{tabular}{cccc}
              $ c_{11}$ & $\frac{1}{{{\tilde{f}}^{24}}_{\; \; \: 4}}$& $ 0$ & $ 0$\\
                $c_{21}$ & 0 & 0 & $ 0$\\
                 0  & 0 & $c_{33}$ & $ c_{34}$\\
                 0  & 0 & $c_{43}$ & $ c_{44}$\\
                 \end{tabular}
                 \right),\;\;\;\;c_{33}c_{44}-c_{34}c_{43}\neq 0;\;\;c_{21}\in\Re-\{0\};\;\;c_{11}\in\Re,
$$\\
where $p=1$ and  ${{\tilde{f}}^{23}}_{\; \; \:
3}={{\tilde{f}}^{24}}_{\; \; \: 4}$,
$$
C_4=\left( \begin{tabular}{cccc}
              $ c_{11}$ & $\frac{p}{{{\tilde{f}}^{24}}_{\; \; \: 4}}$& $ 0$ & $ 0$\\
                $c_{21}$ & 0 & 0 & $ 0$\\
                 0  & 0 & $c_{33}$ & $0$\\
                 0  & 0 & $0$ & $c_{44}$\\
                 \end{tabular}
                 \right),\;\;\;\;c_{21},c_{33},c_{44}\in\Re-\{0\};\;\;c_{11}\in\Re,
$$\\
where $0<|p|\leq 1$  and  ${{\tilde{f}}^{24}}_{\; \; \: 4}=p{{\tilde{f}}^{23}}_{\; \; \: 3}$, and \\
$$
C_5=\left( \begin{tabular}{cccc}
              $ c_{11}$ & $\frac{p}{{{\tilde{f}}^{23}}_{\; \; \: 3}}$& $ 0$ & $ 0$\\
                $c_{21}$ & 0 & 0 & $ 0$\\
                 0  & 0 & $0$ & $ c_{34}$\\
                 0  & 0 & $c_{43}$ & $0$\\
                 \end{tabular}
                 \right),\;\;\;\;c_{21}, c_{34}, c_{43}\in\Re-\{0\};\;\;c_{11}\in\Re,
$$\\
where $0<|p|\leq 1$ and ${{\tilde{f}}^{23}}_{\; \; \:
3}=p{{\tilde{f}}^{24}}_{\; \; \: 4}$.

{$\bullet$} The  solution of $Case \;C$ is isomorphic to the  $(B
\oplus (A_{1,1}+A)), B \oplus A \oplus A_{1,1}$ and $(C^3+A)$ Lie
superalgebras such that isomorphism matrix $C$ between this
solution  and the $(B \oplus (A_{1,1}+A))$ Lie superalgebra is as
follows:
$$
C_1=\left( \begin{tabular}{cccc}
              $ c_{11}$ & $\frac{1}{{{\tilde{f}}^{23}}_{\; \; \: 3}}$& $ 0$ & $ 0$\\
                $2{{\tilde{f}}^{23}}_{\; \; \: 3}c_{43}c_{44}$ & 0 & 0 & $ 0$\\
                 0  & 0 & $c_{33}$ & $ 0$\\
                 0  & 0 & $ c_{43}$ & $ c_{44}$\\
                 \end{tabular}
                 \right),\;\;\;\;c_{33},c_{43},c_{44}\in\Re-\{0\};\;\;c_{11}\in\Re,
$$\\
where ${{\tilde{f}}^{44}}_{\; \; \:1}=\frac{2ic_{43}}{c_{44}}
{{\tilde{f}}^{23}}_{\; \; \: 3}$. The isomorphism matrix  between
this solution  and the $B \oplus A \oplus A_{1,1}$ Lie
superalgebra is  the same as the isomorphism matrix  $C_1$ in the
solution of $Case \;B$ with the same  conditions. Furthermore, the
isomorphism matrix  between this solution  and the $(C^3+A)$ Lie
superalgebra is  the same as the isomorphism matrix  $C_1$ in the
solution of $Case \;A$ with the same conditions.

{$\bullet$} The  solution of $Case \;D$ is isomorphic to the  $(B
\oplus (A_{1,1}+A)), B \oplus A \oplus A_{1,1}$ and $(C^3+A)$ Lie
superalgebras such that isomorphism matrix $C$ between this
solution  and the $(B \oplus (A_{1,1}+A))$ Lie superalgebra is
given by
$$
C_1=\left( \begin{tabular}{cccc}
              $ c_{11}$ & $\frac{1}{{{\tilde{f}}^{24}}_{\; \; \: 4}}$& $ 0$ & $ 0$\\
                $-2{{\tilde{f}}^{24}}_{\; \; \: 4}c_{43}c_{44}$ & 0 & 0 & $ 0$\\
                 0  & 0 & $0$ & $c_{34}$\\
                 0  & 0 & $ c_{43}$ & $ c_{44}$\\
                 \end{tabular}
                 \right),\;\;\;\;c_{34},c_{43},c_{44}\in\Re-\{0\};\;\;c_{11}\in\Re,
$$\\
where ${{\tilde{f}}^{33}}_{\; \; \:1}=-\frac{2ic_{44}}{c_{43}}
{{\tilde{f}}^{24}}_{\; \; \: 4}$. The isomorphism matrix  between
this solution  and the $B \oplus A \oplus A_{1,1}$ Lie
superalgebra is  the same as the isomorphism matrix  $C_2$ in the
solution of $Case \;B$ with the same  conditions. Furthermore, the
isomorphism matrix  between this solution  and the $(C^3+A)$ Lie
superalgebra is  the same as the isomorphism matrix  $C_2$ in the
solution of $Case \;A$ with the same  conditions.

Now using the automorphism supergroup of the $gl(1 | 1)$ and the
method mentioned in \cite{ER1}, one can classify all $gl(1 | 1)$
Lie superbialgebras; the results are presented in table IV.

\section{The $gl(1|1)$ coboundary  Lie superbialgebras}

In this section, we determine which of seventeen  $gl(1|1)$ Lie
superbialgebras of  table IV  are coboundary? Note that here we
work in  nonstandard basis, so we omit the coefficient
$i=\sqrt{-1}$ from all anticommutation relations for the $gl(1|1)$
Lie superalgebra and its dual Lie superalgebras of table IV. In
this regard, we must find $r=r^{ij}X_i \otimes X_j \in {\bf
\mathcal{G}}\otimes{\mathcal{G}}$ such that the super cocommutator
of Lie superbialgebras can be written as in the form (\ref{9}).
Using Eqs. (\ref{9}), (\ref{18}) and (\ref{20}), we have
\cite{ER3}
\begin{equation}
{\tilde{\cal Y}}_i = {{\cal X}_i}^{st} r + (-1)^l\;r {\cal
X}_i,\label{29}
\end{equation}
where index $l$ corresponds to the row of matrix ${\cal X}_i$. Now
using  the above relations, we can find r-matrix of  Lie
superbialgebras and determine which of the presented  Lie
superbialgebras in table IV are coboundary.  We also perform this
work for the dual Lie superbialgebras $(\tilde{ \mathcal{G}},
{\mathcal{G}})$  using the following equations as (\ref{29})
\cite{ER3}
\begin{equation}
{\cal Y}^i = ({\tilde{\cal X}}^i)^{st} \tilde{r} +
(-1)^l\;\tilde{r} {\tilde{\cal X}}^i.\label{30}
\end{equation}

\hspace{-0.10cm}{\footnotesize  TABLE IV.} {\small  Dual Lie
superalgebras to
$gl(1 | 1)$.  }\\
    \begin{tabular}{l l l   p{15mm} }
    \hline\hline
\vspace{-3mm}

&{\footnotesize $ \tilde {\mathcal{G}}$ }& {\footnotesize Non-zero
(anti)
commutation relations}&{\footnotesize Comments}  \smallskip\\
\hline
\smallskip

\vspace{0.5mm}

{\scriptsize Trivial}&{\footnotesize $I_{(2 , 2)}$}& {\footnotesize All of the commutation relations are zero.} \\

\vspace{1mm}

&{\footnotesize ${B \oplus A \oplus
A_{1,1}}_{|}.i$}&{\footnotesize
$[\tilde X^2,\tilde X^3]= \tilde X^3 $}&\\

\vspace{1mm}

&{\footnotesize ${B \oplus A \oplus
A_{1,1}}_{|}.ii$}&{\footnotesize
$[\tilde X^2,\tilde X^4]= \tilde X^4 $}&\\

\vspace{1mm}

&{\footnotesize ${C^2_{p=1}  \oplus
A_{1,1}}_{|}.i$}&{\footnotesize
$[\tilde X^2,\tilde X^3]= \tilde X^3,\;\;\; [\tilde X^2,\tilde X^4]= \tilde X^4 $}&\\

\vspace{1mm}

&{\footnotesize ${C^2_{p=-1}  \oplus
A_{1,1}}_{|}.ii$}&{\footnotesize
$[\tilde X^2,\tilde X^3]= \tilde X^3,\;\;\; [\tilde X^2,\tilde X^4]= -\tilde X^4 $}&\\

\vspace{1mm}

&{\footnotesize ${C^2_{p}  \oplus A_{1,1}}_{|}.i$}&{\footnotesize
$[\tilde X^2,\tilde X^3]= \tilde X^3,\;\;\; [\tilde X^2,\tilde X^4]= p\tilde X^4 $}&{\scriptsize $0 < |p|<1$ }\\

\vspace{1mm}

&{\footnotesize ${C^2_{\frac{1}{p}}  \oplus
{A_{1,1}}_{|}}.ii$}&{\footnotesize
$[\tilde X^2,\tilde X^3]= \tilde X^3,\;\;\; [\tilde X^2,\tilde X^4]=\frac{1}{p} \tilde X^4 $}&{\scriptsize $0 < |p|<1$ }\\

\vspace{-1mm}

{\scriptsize Nontrivial}&{\scriptsize ${\left(B \oplus ( A_{1,1}
\oplus A)\right)}_{\epsilon}.i$}&{\footnotesize
$[\tilde X^2,\tilde X^4]= \epsilon \tilde X^4,\; \;\;[\tilde X^2,\tilde X^3]=
\frac{\epsilon}{2}\tilde X^4,$}\\

\vspace{1mm}

&&{\footnotesize $\{\tilde X^3,\tilde X^3\}= i{\epsilon}\tilde X^1 $}\\

\vspace{-1mm}

&{\scriptsize ${(B \oplus ( A_{1,1} \oplus
A))}_{\epsilon}.ii$}&{\footnotesize
$[\tilde X^2,\tilde X^3]= \epsilon \tilde X^3,\; \;\;[\tilde X^2,\tilde X^4]=
 -\frac{\epsilon}{2}\tilde X^3,$}\\
\vspace{1mm}

&&{\footnotesize $\{\tilde X^4,\tilde X^4\}= i{\epsilon}\tilde X^1$}\\

\vspace{1mm}

&{\footnotesize $(C^3+A)_{\epsilon}.i$}&{\footnotesize
$[\tilde X^2,\tilde X^4]= \frac{\epsilon}{2}\tilde X^3,\; \;\;\{\tilde X^4,\tilde X^4\}
=- i{\epsilon}\tilde X^1 $}&{\scriptsize Nilpotent }\\

\vspace{1mm}

&{\footnotesize $(C^3+A)_{\epsilon}.ii$}&{\footnotesize
$[\tilde X^2,\tilde X^3]= \frac{\epsilon}{2}\tilde X^4,\; \;\;\{\tilde X^3,\tilde X^3\}=
i{\epsilon}\tilde X^1 $}&{\scriptsize Nilpotent }\\

\vspace{-1mm}

&{\footnotesize $(C^2_{-1}+A).i$}&{\scriptsize
$[\tilde X^2,\tilde X^3]= \frac{1}{2}\tilde X^4,\;\;[\tilde X^2,\tilde X^4]=
\frac{1}{2}\tilde X^3,\;\;\{\tilde X^3,\tilde X^3\}=i \tilde X^1,$}\\

\vspace{1mm}

&&{\scriptsize $\{\tilde X^4,\tilde X^4\}=-i \tilde X^1$}\\

\vspace{-1mm}

&{\footnotesize $(C^5_0+A).i$}&{\scriptsize
$[\tilde X^2,\tilde X^3]= \frac{1}{2}\tilde X^4,\;\;[\tilde X^2,\tilde X^4]=
 -\frac{1}{2}\tilde X^3,\;\;\{\tilde X^3,\tilde X^3\}=i \tilde X^1,$} \smallskip\\

&&{\scriptsize $\{\tilde X^4,\tilde X^4\}=i \tilde X^1$} \smallskip\\

\hline\hline
\end{tabular}\\
{\footnotesize $\circ$ ${\epsilon=\pm 1}$.} \vspace{3mm}

\hspace{-0.5cm}{\footnotesize  TABLE V.} {\small  The  coboundary  Lie superbialgebras  ${\mathcal{G}}=gl(1|1)$. }\\
    \begin{tabular}{l l l l l  p{15mm} }
    \hline\hline
\smallskip

&{\footnotesize $({\mathcal{G}}, \tilde {\mathcal{G}})$
}&{\footnotesize  $r$} &{\footnotesize $[[r , r]]$}
\smallskip \\
\hline
\smallskip

\vspace{-0.75mm}

{\footnotesize Triangular }&{\footnotesize $({\mathcal{G}} ,
{C^2_{p=-1} \oplus A_{1,1}}_{|}.ii)$}&
{\footnotesize  $X_1 \wedge X_2$} &{\footnotesize $0$} \\

\vspace{1mm}

{\footnotesize Quasi-triangular }&{\footnotesize $({\mathcal{G}},
{B \oplus A \oplus A_{1,1}}_{|}.i)$}& {\footnotesize  $\frac{1}{2}
X_1 \wedge X_2+\frac{1}{2} X_3 \wedge X_4$}
 &{\footnotesize $-\frac{1}{4} X_2 \wedge X_3 \wedge  X_4$} \\

\vspace{1mm}

&{\footnotesize $({\mathcal{G}} , {B \oplus A \oplus
A_{1,1}}_{|}.ii)$}& {\footnotesize  $-\frac{1}{2} X_1 \wedge
X_2+\frac{1}{2} X_3 \wedge X_4$}
&{\footnotesize $-\frac{1}{4} X_2 \wedge X_3 \wedge  X_4$} \\

\vspace{1mm}

&{\footnotesize $({\mathcal{G}} , {C^2_{p=1}  \oplus
A_{1,1}}_{|}.i)$}&
{\footnotesize  $X_3 \wedge X_4$} &{\footnotesize $- X_2 \wedge X_3 \wedge  X_4$} \\

\vspace{1mm}

&{\footnotesize $({\mathcal{G}} , {C^2_{p}  \oplus
A_{1,1}}_{|}.i)$}& {\footnotesize  $\frac{1-p}{2} X_1 \wedge
X_2+\frac{1+p}{2} X_3 \wedge X_4$} &
{\footnotesize $-\frac{(1+p)^2}{4} X_2 \wedge X_3  \wedge  X_4$}\\

\vspace{-1mm}

&{\footnotesize $({\mathcal{G}} , {C^2_{\frac{1}{p}}  \oplus
{A_{1,1}}_{|}}.ii)$}& {\footnotesize  $\frac{p-1}{2p} X_1 \wedge
X_2+\frac{p+1}{2p} X_3 \wedge X_4$} &  {\footnotesize
$-\frac{(1+p)^2}{4p^2} X_2 \wedge X_3  \wedge  X_4$}
\smallskip\\\hline\hline

\end{tabular}

\vspace{2mm}

Among the  Lie superbialgebras listed in table IV, only six of
them are coboundary and have not coboundary duals. The results are
presented in table V.

\vspace{2mm}
\subsection{Super Poisson
structures on the ${\bf GL( 1 | 1)}$ Lie supergroup }

To obtain the corresponding super Poisson structures on the ${\bf
GL( 1 | 1)}$ Lie supergroup, for the coboundary  Lie
superbialgebras of table V we use the Sklyanin superbracket (the
graded Poisson r-bracket) provided by a given  super
skew-symmetric r-matrix $r=r^{ij}X_i \otimes X_j$
\cite{{N.A},{J.z},{ER3}}. For these reasons  we need the left and
right invariant supervector fields with left (right) derivative on
the $G={\bf GL( 1 | 1)}$ Lie supergroup. To this end, we assume a
convenient parametrization of the ${\bf GL( 1 | 1)}$ Lie
supergroup by means of exponentiation as follows:
\begin{equation}
g = e^{Z^A X_A}= e^{x X_1} e^{y X_2} e^{\psi X_3} e^{\chi
X_4},\;\;\;\;\;\;\;\;g \in G,\label{33}
\end{equation}
which results in
\vspace{-3mm}
$$
{_xX}^{(L,l)}=\frac{{\overrightarrow{\partial}}} {\partial x}-\psi
\frac{{\overrightarrow{\partial}}}{\partial \psi}+\chi
\frac{{\overrightarrow{\partial}}}{\partial \chi},
~~~~~~~{_yX}^{(L,l)}= \frac{{\overrightarrow{\partial}}}{\partial
y},
$$
\vspace{-3mm}
\begin{equation}
{_{\psi}X}^{(L,l)}= -\chi
\frac{{\overrightarrow{\partial}}}{\partial y}-
\frac{{\overrightarrow{\partial}}}{\partial \psi},~~~~~~~
{_{\chi}X}^{(L,l)}=-\frac{{\overrightarrow{\partial}}}{\partial
\chi},~~\label{35}
\end{equation}
\vspace{1mm}
$$
{X_x}^{(L,r)}=\frac{{\overleftarrow{\partial}}}{\partial
x}-\frac{{\overleftarrow{\partial}}}{\partial
\psi}\psi+\frac{{\overleftarrow{\partial}}}{\partial
\chi}\chi,~~~~ ~~~{X_y}^{(L,r)}=
\frac{{\overleftarrow{\partial}}}{\partial y},
$$
\vspace{-3mm}
\begin{equation}
{X_{\psi}}^{(L,r)}= \frac{{\overleftarrow{\partial}}}{\partial y}
\chi - \frac{{\overleftarrow{\partial}}}{\partial \psi},~~~~~~~
{X_{\chi}}^{(L,r)}=-\frac{{\overleftarrow{\partial}}}{\partial
\chi},~~~\label{36}
\end{equation}
\vspace{1mm}
$$
{_xX}^{(R,l)}=\frac{{\overrightarrow{\partial}}} {\partial x},~~~~
~~~~~~~~~~{_yX}^{(R,l)}=
\frac{{\overrightarrow{\partial}}}{\partial y},
$$
\vspace{-3mm}
\begin{equation}
{_{\psi}X}^{(R,l)}=-e^{-x}\frac{{\overrightarrow{\partial}}}{\partial
\psi},~~~~~{_{\chi}X}^{(R,l)}= \psi
e^{x}\frac{{\overrightarrow{\partial}}}{\partial
y}-e^{x}\frac{{\overrightarrow{\partial}}}{\partial
\chi},\label{37}
\end{equation}
\vspace{1mm}
$$
{X_x}^{(R,r)}=\frac{{\overleftarrow{\partial}}}{\partial
x},~~~~~~~~~~~~~~{X_y}^{(R,r)}=\frac{{\overleftarrow{\partial}}}{\partial
y},
$$
\vspace{-3mm}
\begin{equation}
{X_{\psi}}^{(R,r)}= -\frac{{\overleftarrow{\partial}}}{\partial
\psi}e^{-x},~~~~
{X_{\chi}}^{(R,r)}=-\frac{{\overleftarrow{\partial}}}{\partial y}
\psi e^{x}-\frac{{\overleftarrow{\partial}}}{\partial \chi}e^{x},
\label{38}
\end{equation}
where ${_iX}^{(L,l)}({X_i}^{(L,r)})$ and
${_iX}^{(R,l)}({X_i}^{(R,r)})$ stand for left invariant
supervector fields with left(right) derivative and right invariant
supervector fields with left(right) derivative, respectively.
Introducing the generators $Z^A$ representing the supergroup
parameters, one can define for the functions $f(Z^A)$ the graded
Poisson r-bracket as follows \cite{ER3}:
$$
\{f~,~h\}  =
{\{f~,~h\}}^L-{\{f~,~h\}}^R~~~~~~~~~~~~~~~~~~~~~~~~~~~~~~~~~~~~~~~~~~~~~~~~~~
$$
\vspace{-3mm}
$$
~~~~~~~~~~~= \frac{f{\overleftarrow{\partial}}}{\partial
x^{\mu}}\;{^\mu\hspace{-0.5mm}{X^{(L,~r) }_i}}\; r^{ij}
{{_jX}^{(L,~l) }}^\nu \frac{{\overrightarrow{\partial}}h}
{\partial x^{\nu}}- \frac{f{\overleftarrow{\partial}}}{\partial
x^{\mu}}\;{^\mu\hspace{-0.5mm}{X^{(R,\;r) }_i}}\; r^{ij}
{{_jX}^{(R,~l) }}^\nu \frac{{\overrightarrow{\partial}}h}
{\partial x^{\nu}},
$$
\vspace{-3mm}
\begin{equation}
~~~~~~~~~~~~~~~~~~~~~\forall f , h \in C^\infty(G),
\end{equation}
with the following graded anti symmetry property
\begin{equation}
\{f~,~h\}  =  -(-1)^{|f||h|}\{h~,~f\}.
\end{equation}
Now using the  results (\ref{35})-(\ref{38}) and the r-matrices
listed in table V, one can calculate the super Poisson structures
on the ${\bf GL( 1 | 1)}$ Lie supergroup. The results are listed
in the following table (table VI):

\vspace{7mm}

{\scriptsize \hspace{-0.5cm}{\footnotesize  TABLE VI.}
{\footnotesize Poisson superbrackets  the
triangular and quasi-triangular the  coboundary Lie superbialgebras  $\mathcal{G}=gl(1|1)$. }\smallskip\\
    \begin{tabular}{l l l l l l l l l l l p{15mm} }
    \hline\hline
\smallskip

{\scriptsize $({\bf \mathcal{G}} , \tilde {\bf \mathcal{G}})$}
&{\tiny $(\mathcal{G} , C^2_{p=-1}\oplus A_{1,1_{|}}.ii)$}
&\\\hline

\vspace{1mm}

{\footnotesize $\{x,y\}^L$}& {\footnotesize $1$}& {\footnotesize
$\{x,y\}^R$}
&{\footnotesize $1$}& {\footnotesize $\{x,y\}$}&{\footnotesize $0$}&\\

\vspace{1mm}

{\footnotesize $\{x,\psi\}^L$}& {\footnotesize $0$}&{\footnotesize
$\{x,\psi\}^R$}
&{\footnotesize $0$}& {\footnotesize $\{x,\psi\}$}&{\footnotesize $0$}&\\

\vspace{1mm}

{\footnotesize $\{x,\chi\}^L$}& {\footnotesize $0$}&{\footnotesize
$\{x,\chi\}^R$}
&{\footnotesize $0$}& {\footnotesize $\{x,\chi\}$}&{\footnotesize $0$}&\\

\vspace{1mm}

{\footnotesize $\{y,\psi\}^L$}& {\footnotesize
$\psi$}&{\footnotesize $\{y,\psi\}^R$}
&{\footnotesize $0$}& {\footnotesize $\{y,\psi\}$}&{\footnotesize $\psi$}&\\

\vspace{1mm}

{\footnotesize $\{y,\chi\}^L$}& {\footnotesize
$-\chi$}&{\footnotesize $\{y,\chi\}^R$}
&{\footnotesize $0$}& {\footnotesize $\{y,\chi\}$}&{\footnotesize $-\chi$}&\\

\vspace{1mm}

{\footnotesize $\{\psi,\psi\}^L$}& {\footnotesize
$0$}&{\footnotesize $\{\psi,\psi\}^R$}
&{\footnotesize $0$}& {\footnotesize $\{\psi,\psi\}$}&{\footnotesize $0$}&\\

\vspace{1mm}

{\footnotesize $\{\psi,\chi\}^L$}& {\footnotesize
$0$}&{\footnotesize $\{\psi,\chi\}^R$}
&{\footnotesize $0$}& {\footnotesize $\{\psi,\chi\}$}&{\footnotesize $0$}&\\

\vspace{1mm}

{\footnotesize $\{\chi,\chi\}^L$}& {\footnotesize
$0$}&{\footnotesize $\{\chi,\chi\}^R$}
&{\footnotesize $0$}& {\footnotesize $\{\chi,\chi\}$}&{\footnotesize $0$}&\smallskip \\
\hline
{\scriptsize $({\bf \mathcal{G}} , \tilde {\bf \mathcal{G}})$}
&{\tiny $(\mathcal{G} , B \oplus A \oplus A_{1,1_{|}}.i)$}&\
{\tiny $(\mathcal{G} , B \oplus A \oplus A_{1,1_{|}}.ii)$}& {\tiny
$(\mathcal{G} , C^2_{p=1}\oplus A_{1,1_{|}}.i)$}& {\tiny
$(\mathcal{G} , C^2_{p}\oplus A_{1,1_{|}}.i)$}&
{\tiny $(\mathcal{G} , C^2_{\frac{1}{P}}\oplus A_{1,1_{|}}.ii)$}\smallskip\\
\hline

\vspace{1mm}

{\footnotesize $\{x,y\}$}&{\footnotesize $0$}&{\footnotesize $0$}
&{\footnotesize $0$}&{\footnotesize $0$}&{\footnotesize $0$}\\

\vspace{1mm}

{\footnotesize $\{x,\psi\}$}& {\footnotesize $0$}&{\footnotesize
$0$}
&{\footnotesize $0$}&{\footnotesize $0$}&{\footnotesize $0$}\\

 \vspace{1mm}

{\footnotesize $\{x,\chi\}$}& {\footnotesize $0$}&{\footnotesize
$0$}
&{\footnotesize $0$}&{\footnotesize $0$}&{\footnotesize $0$}\\

 \vspace{1mm}

{\footnotesize $\{y,\psi\}$} & {\footnotesize $0$} &
{\footnotesize $-\psi$}&
{\footnotesize $-\psi$}&{\footnotesize $-p \psi$}&{\footnotesize $-\frac{1}{p} \psi$}\\

\vspace{1mm}

{\footnotesize $\{y,\chi\}$} & {\footnotesize $-\chi$} &
{\footnotesize $0$}&
{\footnotesize $-\chi$}&{\footnotesize $-\chi$}&{\footnotesize $-\chi$}\\

\vspace{1mm}

{\footnotesize $\{\psi,\psi\}$} &
{\footnotesize $0$} & {\footnotesize $0$}& {\footnotesize $0$ }&{\footnotesize $0$}&{\footnotesize $0$}\\

\vspace{1mm}

{\footnotesize $\{\psi,\chi\}$} &
{\footnotesize $0$} & {\footnotesize $0$}& {\footnotesize $0$ }&{\footnotesize $0$}&{\footnotesize $0$}\\

\vspace{1mm}

{\footnotesize $\{\chi,\chi\}$} &
{\footnotesize $0$} & {\footnotesize $0$}& {\footnotesize $0$ }&{\footnotesize $0$}&{\footnotesize $0$}\smallskip \\
\hline\hline
\end{tabular}}

\vspace{6mm}
\section{Manin supertriples and Drinfeld superdoubles }

From the Manin supertriples of table IV we see that $(gl{(1 | 1)},
(C^3+A)_{\epsilon_1}.i)$ and $(gl{(1 | 1)},
(C^3+A)_{\epsilon_2}.ii)$ as Lie superalgebras are isomorphic, and
so one can find an isomorphism that preserves also the bilinear
form $<.\;,\;.>$, so that they belong  to the same Drinfeld
superdouble. The isomorphism of Manin supertriples between $(gl{(1
| 1)}, (C^3+A)_{\epsilon_1}.i)$ and $(gl{(1 | 1)},
(C^3+A)_{\epsilon_2}.ii)$  is given by the following
transformations
$$
T^{'}_1  = -T_1 + c T_2 +d T_3,~~~~~~~~~~~~~~~~T^{'}_2 = ab T_2,
$$
$$
T^{'}_3 = -\frac{\epsilon_2 a^2}{\epsilon_1}T_3,~~~~~~~~~T^{'}_4 =
e T_2+f T_3+ \frac{\epsilon_2 a}{\epsilon_1 b}T_4,
$$
$$
  T^{'}_5 = -a T_6,~~~~~~~~~~~~~~~~~~~~~~~~ T^{'}_6=-b T_5,
$$
$$
T^{'}_7 =  -\frac{\epsilon_2 a}{\epsilon_1 }
T_8,~~~~~~~~~~~~~~~T^{'}_8 =- \frac{\epsilon_2 a^2}{\epsilon_1
b}T_7,
$$
where $(T_1,..., T_8)$ are generators of the Manin supertriple
$(gl{(1 | 1)}, (C^3+A)_{\epsilon_1}.i)$ and $(T^{'}_1,...,
T^{'}_8)$ are generators of the Manin supertriples $(gl{(1 | 1)},
(C^3+A)_{\epsilon_2}.ii)$. In same way we get the isomorphism
matrices between all the  Manin supertriples generated by $gl(1|1)
$ (see, Appendix). Furthermore, we deduce the following
theorem.\\
{\bf Theorem 1:} {\it Drinfeld superdoubles generated by the
$gl(1|1)$ Lie superbialgebras of the type $(4 , 4)$  belong to
one of the following 6 classes and allows decomposition into all
Lie superbialgebras listed in the class and their duals.}

\vspace{2mm}

\begin{tabular}{l l l l p{2mm} }

\vspace{2mm}
$\cal{D}$$sd^1_{(4,4)}:$&{\footnotesize $\Big(gl{(1 | 1)}, I_{(2 , 2)}\Big),~
\Big(gl{(1 | 1)}, C^2_{p=-1}\oplus A_{1,1_{|}}.ii\Big),$}&&\\

\vspace{1mm}

$\cal{D}$$sd^{2\;p}_{(4,4)}:$&{\footnotesize $\Big(gl{(1 | 1)},
B \oplus A \oplus A_{1,1_{|}}.i\Big),~ \Big(gl{(1 | 1)}, B \oplus A \oplus A_{1,1_{|}}.ii\Big),
$}&&\\
\vspace{2mm}

&{\footnotesize $\Big(gl{(1| 1)}, C^2_{p=1}\oplus
A_{1,1_{|}}.i\Big),~\Big(gl{(1 | 1)}, C^2_{p}\oplus
A_{1,1_{|}}.i\Big),~\Big(gl{(1 | 1)},
C^2_{\frac{1}{p}}\oplus A_{1,1_{|}}.ii\Big),$}&&\\

\vspace{2mm}

$\cal{D}$$sd^{3\;\epsilon_1, \epsilon_2}_{(4,4)}:$&{\footnotesize $\Big(gl{(1 | 1)},
(B\oplus(A_{1,1}+A))_{\epsilon_1}.i\Big),~\Big(gl{(1 | 1)}, (B\oplus(A_{1,1}+A))_{\epsilon_2}.ii\Big),~{\epsilon_1}, {\epsilon_2} =\pm 1,$}&&\\

\vspace{2mm}

$\cal{D}$$sd^{4\;\epsilon_1, \epsilon_2}_{(4,4)}:$&{\footnotesize $\Big(gl{(1 | 1)},
(C^3+A)_{\epsilon_1}.i\Big),\;\;\Big(gl{(1 | 1)}, (C^3+A)_{\epsilon_2}.ii\Big),~~~{\epsilon_1}, {\epsilon_2} =\pm 1,$}&&\\

\vspace{2mm}

$\cal{D}$$sd^5_{(4,4)}:$&{\footnotesize $\Big(gl{(1 | 1)}, (C^2_{-1}+A).i\Big),$}&&\\

\vspace{2mm}

$\cal{D}$$sd^6_{(4,4)}:$&{\footnotesize $\Big(gl{(1 | 1)}, (C^5_{0}+A).i\Big).$}&&\\

 \end{tabular}

\vspace{3mm}

Drinfeld  superdoubles generated by the $gl(1|1)$  Lie
superbialgebras ${\cal{D}}\;=\;{\cal{D}}_0+{\cal{D}}_1$, are
spanned by the generators $(T_1,..., T_8)$, where $\{T_1, T_2,
T_3, T_4\}$ span the subspace ${\cal{D}}_0$ of grade $0$, and
$\{T_5, T_6, T_7, T_8\}$ span the subsuperspace ${\cal{D}}_1$ of
grade $1$. We give the generators (anti) commutation relations in
table VII. Note that no one of these Drinfeld  superdoubles are
isomorphic to the $osp(2|2)  \cong sl(2|1)$  Lie superalgebra.

\newpage

\hspace{-0.5cm}{\footnotesize  TABLE VII.} {\small  Drinfeld  superdoubles generated by the  Lie superalgebra $ {\mathcal{G}} =gl(1|1)$.  }\\
    \begin{tabular}{l l p{15mm} }
    \hline\hline
{\footnotesize $( \bf {\mathcal{G}} , {\bf {\tilde
{\mathcal{G}}}})$ }& {\footnotesize Non-zero
(anti) commutation relations}  \smallskip\\
\hline
\smallskip

\vspace{-1mm}

{\scriptsize $\left( {\mathcal{G}}, I_{(2,2)}\right)$}&
{\footnotesize $[T_1,T_5]= T_5,\;\;[T_1,T_6]= -T_6,\;\;[T_1,T_7]=
-T_7,\;\;
[T_1,T_8]= T_8,$} \\

\vspace{1mm}

& {\footnotesize $[T_4,T_5]= T_8,\;\;[T_4,T_6]= T_7,\;\;\{T_5,T_6\}= iT_2,\;\;\{T_5,T_7\}= -iT_3,$} \\

\vspace{2mm}

& {\footnotesize $\{T_6,T_8\}= iT_3$} \\


{\scriptsize $\left( {\mathcal{G}}, B \oplus A \oplus
A_{1,1_{|}}.i\right)$}& {\footnotesize $[T_1,T_5]=
T_5,\;\;[T_1,T_6]= -T_6,\;\;[T_1,T_7]= -T_7,\;\;
[T_1,T_8]= T_8,$} \\

\vspace{1mm}

& {\footnotesize $[T_4,T_7]= T_7,\;\;[T_4,T_6]= T_7,\;\;[T_5,T_4]= T_5-T_8,\;\;\{T_5,T_6\}= iT_2,$} \\
\vspace{1mm}

& {\footnotesize $\{T_5,T_7\}= i(T_2-T_3),\;\;\{T_6,T_8\}= iT_3$} \\


{\scriptsize $\left( {\mathcal{G}}, B \oplus A \oplus
A_{1,1_{|}}.ii\right)$}& {\footnotesize $[T_1,T_5]=
T_5,\;\;[T_1,T_6]= -T_6,\;\;[T_1,T_7]= -T_7,\;\;
[T_1,T_8]= T_8,$} \\

\vspace{1mm}

&{\footnotesize $[T_4,T_6]= T_7-T_6,\;\;[T_4,T_8]= T_8,\;\;[T_5,T_4]=-T_8,\;\;\{T_5,T_6\}= iT_2,$} \\
\vspace{1mm}

& {\footnotesize $\{T_5,T_7\}= -iT_3,\;\;\{T_6,T_8\}= i(T_2+T_3)$} \\


{\scriptsize $\left( {\mathcal{G}}, C^2_{p=1}\oplus
A_{1,1_{|}}.i\right)$}& {\footnotesize $[T_1,T_5]=
T_5,\;\;[T_1,T_6]= -T_6,\;\;[T_1,T_7]= -T_7,\;\;
[T_1,T_8]= T_8,$} \\

\vspace{1mm}

& {\footnotesize $[T_4,T_8]= T_8,\;\;[T_4,T_6]= T_7-T_6,\;\;[T_4,T_7]= T_7,\;\;[T_5,T_4]= T_5-T_8,$} \\
\vspace{1mm}

& {\footnotesize $\{T_5,T_6\}= iT_2,\;\;\{T_6,T_8\}= i(T_2+T_3),\;\;\{T_5,T_7\}= i(T_2-T_3)$} \\


{\scriptsize $\left( {\mathcal{G}}, C^2_{p=-1}\oplus
A_{1,1_{|}}.ii\right)$}& {\footnotesize $[T_1,T_5]=
T_5,\;\;[T_1,T_6]= -T_6,\;\;[T_1,T_7]= -T_7,\;\;
[T_1,T_8]= T_8,$} \\

\vspace{1mm}

& {\footnotesize $[T_4,T_8]= -T_8,\;\;[T_4,T_6]= T_6+T_7,\;[T_4,T_7]= T_7,\;\;[T_5,T_4]= T_5-T_8,$} \\

\vspace{1mm}

& {\footnotesize $\{T_5,T_6\}= iT_2,\;\;\{T_6,T_8\}= -i(T_2-T_3),\;\;\{T_5,T_7\}= i(T_2-T_3)$} \\
{\scriptsize $\left( {\mathcal{G}}, C^2_{p}\oplus
A_{1,1_{|}}.i\right)$}& {\footnotesize $[T_1,T_5]=
T_5,\;\;[T_1,T_6]= -T_6,\;\;[T_1,T_7]= -T_7,\;\;
[T_1,T_8]= T_8,$} \\

\vspace{1mm}

& {\footnotesize $[T_4,T_8]= pT_8,\;\;[T_4,T_6]= T_7- pT_6,\;\;[T_4,T_7]= T_7,\;\;[T_5,T_4]= T_5-T_8,$} \\

\vspace{1mm}

& {\footnotesize $\{T_5,T_6\}= iT_2,\;\;\{T_6,T_8\}= i(pT_2+T_3),\;\;\{T_5,T_7\}= i(T_2-T_3)$} \\

{\scriptsize $\left( {\mathcal{G}}, C^2_{\frac{1}{p}}\oplus
A_{1,1_{|}}.ii\right)$}& {\footnotesize $[T_1,T_5]=
T_5,\;\;[T_1,T_6]= -T_6,\;\;[T_1,T_7]= -T_7,\;\;
[T_1,T_8]= T_8,$} \\

\vspace{1mm}

& {\footnotesize $[T_4,T_8]= \frac{1}{p}T_8,\;\;[T_4,T_6]= T_7-\frac{1}{p}T_6,\;\;[T_4,T_7]= T_7,\;\;[T_5,T_4]= T_5-T_8,$} \\

\vspace{1mm}

& {\footnotesize $\{T_5,T_6\}= iT_2,\;\;\{T_6,T_8\}= i(\frac{1}{p}T_2+T_3),\;\;\{T_5,T_7\}= i(T_2-T_3)$} \\


{\scriptsize $\left( {\mathcal{G}}, (B \hspace{-0.5mm} \oplus
\hspace{-0.5mm} (A_{1,1}+ A))_{\epsilon}.i \right)$}&
{\footnotesize $[T_1,T_5]= T_5,\;\;[T_1,T_6]= -T_6,\;\;[T_1,T_7]=
-\epsilon T_5-T_7,\;\;
[T_1,T_8]= T_8,$} \\

\vspace{1mm}

& {\footnotesize $[T_4,T_8]= \epsilon T_8,\;\;[T_4,T_7]=
\frac{\epsilon}{2} T_8,[T_4,T_5]= T_8,\;\;[T_6,T_4]= \epsilon
T_6+\frac{\epsilon}{2}T_5-T_7,$} \\

\vspace{1mm}

& {\footnotesize $\{T_5,T_6\}= iT_2,
\;\;\{T_5,T_7\}= -iT_3,\;\;\{T_6,T_7\}= i\frac{\epsilon}{2}T_2,\{T_6,T_8\}= i(\epsilon T_2+T_3),$} \\
\vspace{1mm}

& {\footnotesize $\{T_7,T_7\}= i\epsilon T_3$} \\


{\scriptsize $\left( {\mathcal{G}}, (B \hspace{-0.5mm}\oplus
\hspace{-0.5mm}(A_{1,1}+ A))_{\epsilon}.ii \right)$}&
 {\footnotesize $[T_1,T_5]= T_5,\;\;[T_1,T_6]= -T_6,\;\;[T_1,T_7]=-T_7 ,\;\;
[T_1,T_8]=-\epsilon T_6+T_8,$} \\

\vspace{1mm}

& {\footnotesize $[T_4,T_8]=-\frac{\epsilon}{2} T_7,\;[T_4,T_7]=
\epsilon T_7,[T_4,T_6]= T_7,\;\;[T_5,T_4]= \epsilon
T_5-\frac{\epsilon}{2}T_6-T_8,
$} \\

\vspace{1mm}

& {\footnotesize $\{T_5,T_6\}= iT_2,\;\;\{T_5,T_7\}= i(\epsilon T_2-T_3),\;\;\{T_5,T_8\}= -i\frac{\epsilon}{2}T_2,\{T_6,T_8\}=iT_3,$} \\

\vspace{1mm}

& {\footnotesize $\{T_8,T_8\}= i\epsilon T_3$}\smallskip \\\hline

\end{tabular}

\newpage

\hspace{-0.5cm}{\footnotesize  TABLE VII. (Cont.)}\\
    \begin{tabular}{l l p{15mm} }
    \hline\hline
{\footnotesize $( \bf {\mathcal{G}} , {\bf {\tilde
{\mathcal{G}}}})$ }& {\footnotesize Non-zero
(anti) commutation relations}  \smallskip\\
\hline
\smallskip

\vspace{-1mm}

{\scriptsize $\left( {\mathcal{G}}, (C^3 +A)_{\epsilon}.i
\right)$}& {\footnotesize $[T_1,T_5]= T_5,\;\;[T_1,T_6]=
-T_6,\;\;[T_1,T_7]= -T_7,\;\;
[T_1,T_8]= T_8+\epsilon T_6,$} \\

\vspace{1mm}

& {\footnotesize $[T_4,T_8]=
\frac{\epsilon}{2}T_7,\;\;[T_6,T_4]=-T_7,\;[T_5,T_4]=
\frac{\epsilon}{2}T_6-T_8,\;\{T_5,T_6\}= iT_2,$} \\
\vspace{1mm}

& {\footnotesize $\{T_5,T_7\}= -iT_3,\;\;\{T_5,T_8\}=
 i\frac{\epsilon}{2}T_2,\;\;\{T_6,T_8\}= iT_3,\;\{T_8,T_8\}= -i\epsilon T_3$} \\


{\scriptsize $\left( {\mathcal{G}}, (C^3 +A)_{\epsilon}.ii
\right)$}& {\footnotesize $[T_1,T_5]= T_5,\;[T_1,T_6]=
-T_6,\;[T_1,T_7]= -T_7-\epsilon T_5,\;\;
[T_1,T_8]= T_8,$} \\

\vspace{1mm}

& {\footnotesize $[T_4,T_7]= \frac{\epsilon}{2}T_8,\;[T_6,T_4]=
\frac{\epsilon}{2}T_5-T_7,\;[T_5,T_4]=-T_8,\;\;\{T_5,T_6\}=
iT_2,$} \\

\vspace{1mm}

& {\footnotesize $\{T_5,T_7\}= -iT_3,\{T_6,T_8\}= iT_3,
\;\;\{T_7,T_7\}= i{\epsilon}T_3,\;\{T_6,T_7\}= i\frac{\epsilon}{2}T_2$} \\

{\scriptsize $\left( {\mathcal{G}}, (C^2_{-1} +A).i \right)$}&
{\footnotesize $[T_1,T_5]= T_5,\;\;[T_1,T_6]=
-T_6,\;\;[T_1,T_7]=-T_5 -T_7,\;\;
[T_1,T_8]=T_6+T_8,$} \\

\vspace{1mm}

& {\footnotesize $[T_4,T_7]= \frac{1}{2}T_8,\;\;[T_4,T_8]=\frac{1}{2} T_7,\;[T_5,T_4]=\frac{1}{2} T_6-T_8,\;\;[T_6,T_4]=\frac{1}{2}T_5 -T_7,$} \\
\vspace{1mm}

& {\footnotesize $\{T_5,T_6\}= iT_2,\;\;\{T_5,T_7\}= -iT_3,\;\;\{T_6,T_8\}= iT_3,\;\;\{T_7,T_7\}= i T_3,$} \\

\vspace{1mm}

& {\footnotesize $\{T_6,T_7\}= \frac{i}{2}T_2,\;\;\{T_8,T_8\}=-i T_3,\;\;\{T_5,T_8\}= \frac{i}{2}T_2$} \\


{\scriptsize $\left({ {\mathcal{G}}, (C^5_{0} +A).i} \right)$}&
{\footnotesize $[T_1,T_5]= T_5,\;\;[T_1,T_6]= -T_6,\;[T_1,T_7]=
-T_5-T_7,\;\;
[T_1,T_8]= T_8- T_6,$} \\

\vspace{1mm}

& {\footnotesize $[T_4,T_7]= \frac{1}{2}T_8,\;[T_4,T_8]=-\frac{1}{2} T_7,\;[T_5,T_4]=-T_8-\frac{1}{2} T_6,\;\;[T_6,T_4]= \frac{1}{2}T_5-T_7,$} \\
\vspace{1mm}

& {\footnotesize $\{T_5,T_6\}= iT_2,\;\;\{T_5,T_7\}= -iT_3,\;\;\{T_6,T_8\}= iT_3,\;\;\{T_7,T_7\}= i T_3,$}\\

\vspace{1mm}

& {\footnotesize $\{T_6,T_7\}= \frac{i}{2}T_2\;\;\{T_8,T_8\}=i T_3,\;\;\{T_5,T_8\}= -\frac{i}{2}T_2$}\smallskip\\

\hline\hline
\end{tabular}

\vspace{4mm}

\section{Integrable system on the $(2|2)-$dimensional homogeneous superspace $OSp(1|2)/U(1)$ }

In this section, we briefly  discuss about the relationship
between the GCYBE and the theory of classical integrable systems
and construct a certain dynamical system on the supersymplectic
supermanifold $OSp(1|2)/U(1)$.  We begin by recalling the notion
of a completely integrable Hamiltonian system. A  Hamiltonian
system is modeled by a Poisson supermanifold ${\cal M}$ ( $dim
{\cal M}$ is denote by $(d_B|d_F)$ where $d_B(d_F)$ is the
dimension of the bosonic (fermionic) part.) and a Hamiltonian
${\cal H} \in C^{\infty}({\cal M})$, such that its time-evolution
is given by
\begin{equation}
X_{\cal H}\;=\;\{{\cal H}\;,\; f \},\label{integrable:1}
\end{equation}
where $X_{\cal H}$  is  the Hamiltonian supervector field on $\cal
M$ corresponding to ${\cal H}$ and  $f \in C^{\infty}({\cal M})$.
Whenever $m(t)$ be any integral curve of $X_{\cal H}$, then
$\left( df/dt\right)(m(t))\\=\{{\cal H}\;,\; f \}(m(t))$. In
particular, an observable $f$ is conserved, or a constant of the
motion, if $\{{\cal H}\;,\; f \} = 0$, or more generally, two
observables $f_1$ and $f_2$ on a Poisson supermanifold $\cal M$
are in involution if $\{f_1, f_2\} = 0$. Note that every integral
curve $m(t)$ of $X_{\cal H}$  lies entirely in some
supersymplectic leaf of $\cal M$. Hence, we might as well assume
that $\cal M$ is a supersymplectic supermanifold.\\
{\bf Definition 6:} {\it The dynamical system defined on a
$(d_B|d_F)$-dimensional supersymplectic supermanifold $\cal M$
(here $d_B$ and $d_F$ are even numbers) by a Hamiltonian ${\cal
H}\in C^{\infty}({\cal M})$ is completely integrable if there
exist $n=\frac{1}{2} dim {\cal M}$ independent conserved
quantities $f_1,...f_n\in C^{\infty}({\cal M})$ in involution.}

We now  give a general procedure for constructing completely
integrable Hamiltonian systems starting from a classical r-matrix.
Let $\cal M$ be a supermanifold with a non-degenerate
supersymplectic form $\omega$, and $x^{\mu} (\mu=1,...,dim {\cal
M})$ be the local coordinates of ${\cal M}$. Consider Poisson
superbracket structure on ${\cal M}$ for arbitrary functions
$f(x^{\mu})$ and $g(x^{\nu})$ as \cite{Gould}
\begin{equation}
\{ f\;,\;g\}\;=\;\  \frac{f{\overleftarrow{\partial}}}{\partial
x^{\mu}}\;\omega^{\mu\nu}\; \frac{{\overrightarrow{\partial}}g}
{\partial
x^{\nu}}=(-1)^{\nu(\mu+|f|)}\;\omega^{\mu\nu}\frac{{\overrightarrow{\partial}}f}
{\partial x^{\mu}}\frac{{\overrightarrow{\partial}}g} {\partial
x^{\nu}},\label{integrable:2}
\end{equation}
where $\omega^{\mu\nu}$ is the superinverse of the
$\omega_{\mu\nu}$, such that
$\omega^{\mu\nu}=-(-1)^{\mu\nu}\omega^{\nu\mu}$. Then, by
introducing dynamical variables $S_i(x^{\mu}), \;i=1,...,dim
{{\mathcal{G}}}$, as
\begin{equation}
\{ S_i\;,\;S_j\}\;=\;f_{ij}^{\;\;k}\;S_k,\label{integrable:3}
\end{equation}
and  defining the  Lie superalgebra-valued function
\begin{equation}
Q(x)\;=\;(-1)^j S_i(x)\; r^{ij}\;X_j,\label{integrable:4}
\end{equation}
and  using the fact that $r$ satisfies the GCYBE, it is concluded
that \cite{Gould}
\begin{equation}
\{ Str[(Q(x))^n ]\;,\;Str[(Q(xy))^{m} ]\}\;=\;0,~~~~~~~~~~0<m,\;n
\in \mathbb{Z^+}.\label{integrable:5}
\end{equation}
Therefore, the coefficients
\begin{equation}
I_k(x)= Str\left[(Q(x))^k \right],~~~~~~~~~~0<k \in
\mathbb{Z^+},\label{integrable:5.1}
\end{equation}
are regarded as constants of motion of a certain dynamical system.
In the following by using the above statements we construct an
integrable system on a supersymplectic homogeneous superspace. It
has been shown that $OSp(1|2)$ coherent states are parametrized by
points of a supersymplectic supermanifold, namely the homogeneous
superspace $OSp(1|2)/U(1)$, such that $OSp(1|2)/U(1)$ is a
supercoadjoint orbit of $OSp(1|2)$, i.e., a superunit disc
\cite{Amine}. This superunit disc can be equipped with a
supersymplectic even two-form $\omega$ that is given by
\cite{Amine}
$$
\omega =-2i\tau \left[1+\frac{1}{2}
\theta^{^{\#}}\theta{(1+|z|^2)\over(1-|z|^2)}\right]\frac{dz
\wedge d \overline{z}}{(1-|z|^2)^2}+i\tau \frac{d\theta \wedge d
\theta^{^{\#}}}{(1-|z|^2)}~~~~~~~~~~~~~~~~~~
$$
\vspace{-3mm}
\begin{equation}
~~~~~~~~~~~+i\tau \theta \overline{z} \frac{dz \wedge d
\theta^{^{\#}}}{(1-|z|^2)^2}-i\tau z \theta^{^{\#}} \frac{d\theta
\wedge d \overline{z}}{(1-|z|^2)^2}.\label{integrable:6}
\end{equation}
where $\tau$ corresponds to each irreducible representation  of
$OSp(1|2)$.  Here $z$ is a complex number  and $\theta$ is an odd
coordinate, i.e. an anticommuting Grassmann number. Note that
$\overline{z}$ is the usual complex conjugate of $z$ while
$\theta^{\#}$ is the so-called adjoint of $\theta$ (one can  write
them in terms of "real" coordinates $X, Y$ and ${\psi}^+,
{\psi}^-$ as $z=X+iY,\; \overline{z}=X-iY$ and
$\theta={\psi}^{+}+i{\psi}^-,
\;\theta^{\#}=-i({\psi}^{+}-i{\psi}^{-})$ \cite{Amine}.).

Now, using the definition of two-form \cite{D} as
$\omega=\frac{(-1)^{\mu\nu}}{2}\;\omega_{\mu\nu} \;dx^{\mu} \wedge
dx^{\nu}$, one can find the superinverse of the $\omega_{\mu\nu}$
as follows:
\begin{equation}
{\omega^{\mu \nu}=\frac{1}{4i\tau}\left(
\begin{tabular}{cccc}
              $ 0$ & $\frac{\left (2B-\theta^{^{\#}}\theta \right)B}{2}$& $ 0$ & $-{z \theta^{^{\#}}B}$\\
                $-\frac{\left (2B-\theta^{^{\#}}\theta\right)B}{2}$ & 0 &
                 ${ \overline{z} \theta B}$ & 0\\
                 0  & $-{ \overline{z} \theta B}$& 0 & ${2 B- z \overline{z} \theta^{^{\#}}\theta}$\\
                  ${z \theta^{^{\#}}B}$  & 0 & ${2 B- z \overline{z} \theta^{^{\#}}\theta}$ & 0\\
                 \end{tabular}
                 \right).}~~~~~~~~~~\label{integrable:7}
\end{equation}\\

Thus, by taking the structure constants of the $gl(1|1)$
(\ref{23}) and employing  the relations (\ref{integrable:2}) and
(\ref{integrable:3}) we then obtain the following system of PDEs
{\footnotesize
$$
\frac{B(2B-\theta^{^{\#}}\theta)}{2}\left[\frac{{\partial }
S_3}{\partial z}\frac{{\partial } S_4}{\partial
\overline{z}}-\frac{{\partial } S_3}{\partial
\overline{z}}\frac{{\partial } S_4}{\partial z }\right]+(B z
\theta^{^{\#}})\left[\frac{{\partial } S_3}{\partial
z}\frac{{\partial } S_4}{\partial \theta^{^{\#}}}-\frac{{\partial
} S_3}{\partial \theta^{^{\#}}}\frac{{\partial } S_4}{\partial z
}\right]~~~~~~~~~~~~~~~~
$$}
\vspace{-2mm} {\footnotesize
\begin{equation}
~~~~~~~- (B \overline{z} \theta )\left[\frac{{\partial }
S_3}{\partial \overline{z}}\frac{{\partial } S_4}{\partial
\theta}-\frac{{\partial } S_3}{\partial \theta}\frac{{\partial }
S_4}{\partial \overline{z} }\right]+(2B +|z|^2 \theta
\theta^{^{\#}} )\left[\frac{{\partial } S_3}{\partial \theta
}\frac{{\partial } S_4}{\partial \theta^{^{\#}}}+\frac{{\partial }
S_3}{\partial \theta^{^{\#}}}\frac{{\partial } S_4}{\partial
\theta }\right]=S_2,~~~~~\label{integrable:8}
\end{equation}}
{\footnotesize
$$
\frac{B(2B-\theta^{^{\#}}\theta)}{2}\left[\frac{{\partial }
S_2}{\partial z}\frac{{\partial } S_i}{\partial
\overline{z}}-\frac{{\partial } S_2}{\partial
\overline{z}}\frac{{\partial } S_i}{\partial z }\right]-(B z
\theta^{^{\#}})\left[\frac{{\partial } S_2}{\partial
z}\frac{{\partial } S_i}{\partial \theta^{^{\#}}}-\frac{{\partial
} S_2}{\partial \theta^{^{\#}}}\frac{{\partial } S_i}{\partial z
}\right]~~~~~~~~~~~~~~
$$}
\vspace{-2mm} {\footnotesize
\begin{equation}
~~~~~~~~+ (B \overline{z} \theta )\left[\frac{{\partial }
S_2}{\partial \overline{z}}\frac{{\partial } S_i}{\partial
\theta}-\frac{{\partial } S_2}{\partial \theta}\frac{{\partial }
S_i}{\partial \overline{z} }\right]-(2B +|z|^2 \theta
\theta^{^{\#}} )\left[\frac{{\partial } S_2}{\partial \theta
}\frac{{\partial } S_i}{\partial \theta^{^{\#}}}+\frac{{\partial }
S_2}{\partial \theta^{^{\#}}}\frac{{\partial } S_i}{\partial
\theta }\right]=0,\label{integrable:9}
\end{equation}}

\newpage

{\footnotesize
$$
\frac{B(2B-\theta^{^{\#}}\theta)}{2}\left(\frac{{\partial }
S_i}{\partial z}\frac{{\partial } S_i}{\partial
\overline{z}}\right)+(B z \theta^{^{\#}})\left(\frac{{\partial }
S_i}{\partial z}\frac{{\partial } S_i}{\partial
\theta^{^{\#}}}\right) - (B \overline{z} \theta
)\left(\frac{{\partial } S_i}{\partial
\overline{z}}\frac{{\partial } S_i}{\partial \theta}\right)
$$}
\vspace{-2mm} {\footnotesize
\begin{equation}
~~~~~~~~~~~~~~~~~~~~~~~~~~~~~~~+(2B +|z|^2 \theta
\theta^{^{\#}})\left(\frac{{\partial } S_i}{\partial \theta
}\frac{{\partial } S_i}{\partial
\theta^{^{\#}}}\right)=0,\label{integrable:10}
\end{equation}}
{\footnotesize
$$
\frac{B(2B-\theta^{^{\#}}\theta)}{2}\left[\frac{{\partial }
S_1}{\partial z}\frac{{\partial } S_j}{\partial
\overline{z}}-\frac{{\partial } S_1}{\partial
\overline{z}}\frac{{\partial } S_j}{\partial z }\right]-(B z
\theta^{^{\#}})\left[\frac{{\partial } S_1}{\partial
z}\frac{{\partial } S_j}{\partial \theta^{^{\#}}}-\frac{{\partial
} S_1}{\partial \theta^{^{\#}}}\frac{{\partial } S_j}{\partial z
}\right]~~~~~~~~~~~~~~~~~~~~~~~~~~~~~~~
$$}
\vspace{-2mm} {\footnotesize
\begin{equation}
+(B \overline{z} \theta )\left[\frac{{\partial } S_1}{\partial
\overline{z}}\frac{{\partial } S_j}{\partial
\theta}-\frac{{\partial } S_1}{\partial \theta}\frac{{\partial }
S_j}{\partial \overline{z} }\right]-(2B +|z|^2 \theta
\theta^{^{\#}} )\left[\frac{{\partial } S_1}{\partial \theta
}\frac{{\partial } S_j}{\partial \theta^{^{\#}}}+\frac{{\partial }
S_1}{\partial \theta^{^{\#}}}\frac{{\partial } S_j}{\partial
\theta }\right]=\left\{\begin{array}{lll}0&{\rm } \;j =2\\S_3&{\rm } \;j =3\\
-S_4&{\rm } \;j =4\\\end{array} \right.,\label{integrable:11}
\end{equation}}
where $i=3, 4;\;B=1-|z|^2$ and $\frac{{\partial }}{{\partial
x^{\mu}}}$ stands for left derivative with respect to $x^{\mu}$.
A special class of  solutions for the above equations is given by
{\footnotesize
\begin{equation}
S_1=\frac{2i\tau \theta^{^{\#}} \theta}{(1-|z|^2)},\;\;
S_2=\frac{\alpha_0\beta_0}{2i\tau} \left(1-|z|^2+\frac{1}{2}|z|^2
\theta \theta^{^{\#}} \right),\;\; S_3=\alpha_0
\theta^{^{\#}},\;\;S_4=\beta_0 \theta, \label{integrable:16}
\end{equation}}
where $\alpha_0$ and $\beta_0$ are even constants, such that
$\alpha_0\beta_0=i\mathcal{C}$ and $\mathcal{C}$ is a real
constant. On the other hand, among the coboundary Lie
superbialgebras listed in table V, only the $(gl(1|1) ,
{C^2_{p=-1}  \oplus A_{1,1}}_{|}.ii)$ satisfies the GCYBE,
therefore, using the classical $r$-matrix concerning this   Lie
superbialgebra (table V), the relation (\ref{integrable:4}) takes
on the following form
\begin{equation}
Q\;=\; S_1 X_2-S_2 X_1. \label{integrable:17}
\end{equation}
It is easy to verify that the following supermatrices provide a
$(2|2)$-dimensional representation of $gl(1|1)$
$$
X_1=\left( \begin{tabular}{cccc}
              $n$ & 0 & $ 0 $ & 0\\
              $0$ & n & $ 0 $ & 0\\
              $0$ & 0 & $ n-1 $ & 0\\
              $1$ & 0 & $ -1 $ & $n+1$\\
                 \end{tabular}
                 \right),  ~~~~ \;\;\;\;\; X_2=\left( \begin{tabular}{cccc}
              $0$ & 0 & $ 0 $ & 0\\
              $-ie$ & ie & $ 0 $ & 0\\
              $0$ & 0 & $ 0 $ & 0\\
              $ie$ & 0 & $ -\frac{ie}{2} $ & $ie$\\
                 \end{tabular}
                 \right),~~~~~~~~
$$
\begin{equation}
 X_3=\left( \begin{tabular}{cccc}
              $0$ & 0 & $ 0 $ & 0\\
              $0$ & 0 & $ 0 $ & 0\\
              $0$ & 0 & $ 0$ & 0\\
              $1$ & -1 & $ 0 $ & 0\\
                 \end{tabular}
                 \right),\;\;\;~~~~\;\;\;\;\;X_4=\left( \begin{tabular}{cccc}
              $0$ & 0 & $ 0 $ & 0\\
              $-ie$ & 0 & $ \frac{ie}{2} $ & -ie\\
              $0$ & 0 & $ 0 $ & 0\\
              $0$ & 0 & $ 0 $ & 0\\
                 \end{tabular}
                 \right), \label{integrable:18}
\end{equation}
where $e$ and $n$ are real constants. Thus, substituting the above
relations into (\ref{integrable:17}) and with the help of
(\ref{integrable:5.1}), we obtain the constants of motion as
follows:
\begin{equation}
I_2(x^\mu)=-4e\mathcal{C} {\psi}^{+} {\psi}^{-}
-\frac{\mathcal{C}^2}{2 {\tau}^2}
(1-X^2-Y^2)\left[1-X^2-Y^2-2(X^2+Y^2){\psi}^{+}
{\psi}^{-}\right],~~~~~~~~~~~~~~~~~~~~~~~~ \label{integrable:19}
\end{equation}
$$
I_3(x^\mu)=\frac{3e(2n+1)\mathcal{C}^2}{{\tau}}
(1-X^2-Y^2){\psi}^{+}
{\psi}^{-}~~~~~~~~~~~~~~~~~~~~~~~~~~~~~~~~~~~~~
$$
\vspace{-3mm}
\begin{equation}
~~~~~~~~~~~~~~~~~~~~~~~~~ +\frac{3n\mathcal{C}^3}{4 {\tau}^3}
(1-X^2-Y^2)^2\left[1-X^2-Y^2-3(X^2+Y^2){\psi}^{+}
{\psi}^{-}\right], \label{integrable:20}
\end{equation}
where $X, Y, {\psi}^{+}$ and ${\psi}^{-}$ are the superphase space
coordinates.  One can consider $I_2(x^\mu)$ as the Hamiltonian of
the system and $I_3(x^\mu)$ as other constant of motion or vice
versa. Note that $I_1(x^\mu)$ is also vanished.

\section{The quantization of  $gl(1|1)$ Lie superalgebra}

In this section we quantize $gl(1|1)$ Lie superalgebra by use of
the Lyakhovsky and Mudrov formalism \cite{Lyakhovsky}. For this
purpose and self-containing of the paper we first review the main
result of this formalism as a following proposition; then we use
this method in order to build up the Hopf superalgebras related to
all $gl(1|1)$ coboundary Lie superbialgebras of table
V.\\

{\bf Proposition 2} \cite{Lyakhovsky}: {\it Let $\{1, H_1, \cdots
, H_n,  X_1, \cdots , X_m\}$ be a basis of an associative algebra
${\cal A}$ over $C$ such that  satisfying the following condition

\begin{equation}
[H_i , H_j]\;=\;0,\qquad i, j\;=\;1, \cdots ,n.\label{Quantiz.1}
\end{equation}
In addition, let  $\mu_i$ and $\nu_j ~(i, j = 1, \cdots ,n)$ be a
set of $m\times m$ complex matrices such that they are commute
with together.

Let ${\overrightarrow {X}}$ be a column vector with component $X_l
(l= 1, \cdots ,m)$, the {\it coproduct, counit} and {\it antipode}
are defined as
\begin{equation}
\triangle(1)\;=\;1 \otimes 1, \qquad \bigtriangleup
(H_i)\;=\;I\otimes H_i + H_i \otimes I,\label{Quantiz.2}
\end{equation}
\begin{equation}
\triangle({\overrightarrow {X}})\;=\;\exp(\sum^{n}_{i=1}
{\mu_i}{H_i}) \dot{\otimes} {\overrightarrow {X}} + \sigma
\Big(\exp(\sum^{n}_{i=1} {\nu_i}{H_i}) \dot{\otimes}
{\overrightarrow {X}}\Big),\label{Quantiz.3}
\end{equation}
\begin{equation}
\epsilon(1)\;=\;1, \qquad  \epsilon(H_i)\;=\; \epsilon
(X_l)\;=\;0, ~~~ i \;=\;1, \cdots ,n, ~~ l\;=\;1, \cdots
,m,\label{Quantiz.4}
\end{equation}
\vspace{-1mm}
\begin{equation}
S(H_i)\;=\;-H_i \qquad  S(X)\;=\; -e^{{\mu_i}{H_i}} X
e^{{\nu_i}{H_i}}, \qquad S(1)\;=\;1,\label{Quantiz.5}
\end{equation}
then  $({{\cal A}}, \Delta, \epsilon, {\cal S})$  endow with a
Hopf algebra structure.}\\
\smallskip
The cocommutator $\delta$ can be written in the following matrix
form
\begin{equation}
\delta ({\overrightarrow
{X}})\;=\;\bigtriangleup_{(1)}({\overrightarrow {X}})-\sigma \circ
\bigtriangleup_{(1)}({\overrightarrow {X}}),\label{Quantiz.6}
\end{equation}
where $\bigtriangleup_{(1)}({\overrightarrow {X}})$ is the first
order of (\ref{Quantiz.3}), i.e.,
\begin{equation}
\bigtriangleup_{(1)}({\overrightarrow {X}})\;=\;(\sum^{n}_{i=1}
{\mu_i}{H_i}) \dot{\otimes} {\overrightarrow {X}} + \sigma
(\sum^{n}_{i=1} {\nu_i}{H_i} \dot{\otimes} {\overrightarrow
{X}}).\label{Quantiz.7}
\end{equation}
In this formalism, elements $H_i$ are called  primitive
generators. These elements must be chosen  such that $\delta
({\overrightarrow {X}}_i)$ does not contain terms of the form $H_i
\wedge H_j$. We note that the same cocommutator (\ref{Quantiz.6})
can be obtained from different choices of the matrices $\mu_{i}$
and $\nu_{j}$, i.e., the different sets of matrices  lead to right
quantization of $U({\mathcal{G}})$. In this way one  can choose
$\mu _i\;=0$ as a representative of all these quantizations and
obtain
\begin{equation}
\delta ({\overrightarrow {X}})\;=\;-(\sum^{n}_{i=1} {\nu_i}{H_i})
\dot{\wedge} {\overrightarrow {X}}\;=\;-(\sum^{n}_{i=1}
{\nu_i}{H_i}) \dot{\otimes} {\overrightarrow {X}} + \sigma
(\sum^{n}_{i=1} {\nu_i}{H_i} \dot{\otimes} {\overrightarrow
{X}}).\label{Quantiz.8}
\end{equation}
When the algebra ${\cal A}$ is a Lie algebra ${\mathcal{G}}$, then
$\{ H_i \}$ generate an Abelian Lie subalgebra and with this
condition the deformed commutation relations in
$U_h({\mathcal{G}})$ are given by \cite{Lyakhovsky}
\begin{equation}
[X_l , X_p]\;=\;[X_l , X_p]_{\circ} + \phi_{lp}(\mu_i, \nu_j,
H_k),\label{Quantiz.9}
\end{equation}
where $[X_l , X_p]_{\circ}$ is the classical commutation relation
and the deforming functions $\phi_{lp}$  are the power series of
$H_k$'s.  Note that after determining of $\phi_{lk}$, the Jacobi
identity for (\ref{Quantiz.9})  must be checked.

The above formalism was presented for the Lie algebras, one can
use this formalism for Lie superalgebras by keeping that the
graded tensor product law must be taken into account \cite{kulish}
\begin{equation}
(F \otimes G)_{ij;kl}\;=\;(-1)^{j(i+k)}\;
F_{ik}G_{jl}.\label{Quantiz.10}
\end{equation}
This quantization procedure can be applied to the coboundary Lie
superbialgebras of table V to quantize  the $gl(1|1)$ Lie
superalgebra. To this end, we use the following $2 \times 2$ real
matrix representation of the basis of $gl(1|1)$ {\footnotesize
\begin{equation} \bf X_1=\left( \begin{tabular}{cc}
              $ \frac{1}{2}$ &$0$ \\
              $0$ & $ -\frac{1}{2}$  \\
              \end{tabular}
              \right),~~\bf X_2=\left( \begin{tabular}{cc}
              $1$ &$0$ \\
              $0$ & $1$  \\
              \end{tabular}
              \right),~\bf X_3=\left( \begin{tabular}{cc}
              $0$ &$1$ \\
              $0$ & $0$  \\
              \end{tabular}
              \right),~\bf X_4=\left( \begin{tabular}{cc}
              $0$ &$0$ \\
              $1$ & $ 0$  \\
              \end{tabular}
              \right).   \label{Quantiz.11}
\end{equation}}
The supercoproducts for the corresponding quantum universal
enveloping superalgebra of the coboundary Lie superbialgebras of
table V are written down as four propositions. We denote each
multiparametric quantum Hopf superalgebra by ${{
U}_h}^{\hspace{-1mm}({ \tilde \mathcal{G}})}( gl(1|1) )$. Hence,
quantum deformations for the $gl(1|1)$ Lie superalgebra are given
by the following statements.

{\bf Proposition 3:} The  supercoproduct $\Delta$, supercounit
$\epsilon$,  superantipode $S$
\begin{equation}
\Delta(X_i)\;=1 \otimes X_i + X_i \otimes 1,~~~~~~~~
 \label{Quantiz.12}
\end{equation}
\begin{equation}
\Delta(X_3)\;=1 \otimes X_3 + X_3 \otimes e^{-hX_2},~~
 \label{Quantiz.13}
\end{equation}
\begin{equation}
~~~~~~~~~~~~~~\epsilon(1)\;=1,~~~ \epsilon({\bf X})\;=0,~~~{\bf
X}\in \{X_1, X_2,  X_3, X_4\},
 \label{Quantiz.14}
\end{equation}
\begin{equation}
S(X_i)\;=\;- X_i,~~~~~~~~~~~~~~~~~~~~~~
 \label{Quantiz.15}
\end{equation}
\begin{equation}
S(X_3)\;=\;- X_3 e^{hX_2},~~~~i=1, 2, 4
 \label{Quantiz.16}
\end{equation}
and the non-zero super (anti)commutation relations
\begin{equation}
[X_1 , X_3]\;=\;X_3,~~~~~~[X_1 , X_4]\;=\;-X_4,~~~~~~\{X_3 ,
X_4\}\;=\;\frac{1-e^{-hX_2}}{h}.
 \label{Quantiz.17}
\end{equation}
determine a Hopf superalgebra denoted by  ${{
U}_h}^{\hspace{-1mm}( {B \oplus A \oplus A_{1,1}}_{|}.i)}( gl(1|1)
)$ which quantizes the quasi-triangular Lie superbialgebra
$(gl(1|1) , {B \oplus A \oplus A_{1,1}}_{|}.i )$.

{\bf Proposition 4:} The Hopf superalgebra denoted by ${{
U}_h}^{\hspace{-1mm}( {B \oplus A \oplus A_{1,1}}_{|}.ii)}(
gl(1|1) )$ which quantizes the quasi-triangular Lie superbialgebra
$(gl(1|1) , {B \oplus A \oplus A_{1,1}}_{|}.ii )$ has supercounit
(\ref{Quantiz.14}) and the following supercoproduct and
superantipode
\begin{equation}
\Delta(X_i)\;=1 \otimes X_i + X_i \otimes 1,~~~~~~~~
 \label{Quantiz.18}
\end{equation}
\begin{equation}
\Delta(X_4)\;=1 \otimes X_4 + X_4 \otimes e^{-hX_2},~~
 \label{Quantiz.19}
\end{equation}
\begin{equation}
S(X_i)\;=\;- X_i,~~~~~~~~~~~~~~~~~~~~~~
 \label{Quantiz.20}
\end{equation}
\begin{equation}
S(X_4)\;=\;- X_4 e^{hX_2},~~~~i=1, 2, 3
 \label{Quantiz.21}
\end{equation}
with the same non-zero super (anti)commutation relations deduced
in (\ref{Quantiz.17}).

{\bf Proposition 5:} The  quantum superalgebra which quantizes the
triangular Lie superbialgebra $(gl(1|1) , {C_{p=-1}^2 \oplus
A_{1,1}}_{|}.ii )$ has Hopf structure denoted by ${{
U}_h}^{\hspace{-1mm}({C_{p=-1}^2 \oplus A_{1,1}}_{|}.ii)}( gl(1|1)
)$ and is characterized by the supercounit (\ref{Quantiz.14}) and
the following  supercoproduct and  superantipode
\begin{equation}
\Delta(X_i)\;=1 \otimes X_i + X_i \otimes 1,~~~~~~~~
 \label{Quantiz.22}
\end{equation}
\begin{equation}
\Delta(X_3)\;=1 \otimes X_3 + X_3 \otimes e^{-hX_2},~
 \label{Quantiz.23}
\end{equation}
\begin{equation}
\Delta(X_4)\;=1 \otimes X_4 + X_4 \otimes e^{hX_2},~
 \label{Quantiz.23}
\end{equation}
\begin{equation}
S(X_i)\;=\;- X_i,~~~~~~~~~~~~~~~~~~~
 \label{Quantiz.25}
\end{equation}
\begin{equation}
S(X_3)\;=\;- X_3 e^{hX_2},~~~~~~~~~~~~~
 \label{Quantiz.26}
\end{equation}
\begin{equation}
~~S(X_4)\;=\;- X_4 e^{-hX_2},~~~~i=1, 2
 \label{Quantiz.26}
\end{equation}
together with the non-zero super (anti)commutation relations
\begin{equation}
[X_1 , X_3]\;=\;X_3,~~~~~~~~[X_1 , X_4]\;=\;-X_4,~~~~~~~~\{X_3 ,
X_4\}\;=\;(1-h)X_2.
 \label{Quantiz.27}
\end{equation}

{\bf Proposition 6:} The  Hopf superalgebras denoted by ${{
U}_h}^{\hspace{-1mm}({C_{\lambda}^2 \oplus A_{1,1}})}( gl(1|1) )$
(where $\lambda = 1, p, \frac{1}{p}$) which quantize the
quasi-triangular Lie superbialgebras $\left(gl(1|1) ~, ~\tilde
{\mathcal{G}}={C_{p=1}^2 \oplus A_{1,1}}_{|}.i, {C_{p}^2 \oplus
A_{1,1}}_{|}.i, {C_{\frac{1}{p}}^2 \oplus A_{1,1}}_{|}.ii\right)$
have supercounit (\ref{Quantiz.14}), supercoproduct, superantipode
\begin{equation}
\Delta(X_i)\;=1 \otimes X_i + X_i \otimes 1,~~~~~~~~~~
 \label{Quantiz.22}
\end{equation}
\begin{equation}
\Delta(X_3)\;=1 \otimes X_3 + X_3 \otimes e^{-hX_2},~~~
 \label{Quantiz.23}
\end{equation}
\begin{equation}
\Delta(X_4)\;=1 \otimes X_4 + X_4 \otimes e^{-\lambda h X_2},~~
 \label{Quantiz.23}
\end{equation}
\begin{equation}
S(X_i)\;=\;- X_i,~~~~~~~~~~~~~~~~~~~~~~~
 \label{Quantiz.25}
\end{equation}
\begin{equation}
S(X_3)\;=\;- X_3 e^{hX_2},~~~~~~~~~~~~~~~~~
 \label{Quantiz.26}
\end{equation}
\begin{equation}
S(X_4)\;=\;- X_4 e^{\lambda hX_2},~~~~i=1, 2~~
 \label{Quantiz.26}
\end{equation}
and the non-zero super (anti)commutation relations
\begin{equation}
[X_1 , X_3]\;=\;X_3,~~~~~~[X_1 , X_4]\;=\;-X_4,~~~~~~\{X_3 ,
X_4\}\;=\;\frac{1-e^{-(1+\lambda) h X_2}}{(1+\lambda)h}.
 \label{Quantiz.27}
\end{equation}
In the following, we obtain  the Hopf structure of the associated
quantum supergroup $GL_h(1|1)$ by using the classical r-matrix
related to the triangular Lie superbialgebra $(gl(1|1) ,
{C_{p=-1}^2 \oplus A_{1,1}}_{|}.ii )$. Quantum R-matrix satisfying
graded QYB equation can be calculated as
$$
R\;=\;e^{hr}\;=\;e^{h(X_1 \otimes X_2-
 X_2 \otimes X_1)}\hspace{3.5cm}~~~~~~~~~~~~~~~~~~~~~~~~~~~~~~~~~~~~~~~~~
$$
\begin{equation}
~~~~~~~~~~~~~~~=2(1-\cosh h)X_1 \otimes X_1+\frac{1}{2}(1+\cosh
h)X_2 \otimes X_2+\sinh h X_1 \wedge X_2,
 \label{Quantiz.28}
\end{equation}
using Eqs. (\ref{Quantiz.10}) and (\ref{Quantiz.11}) one can
obtain\footnote{For Hopf superalgebra ${{
U}_h}^{\hspace{-1mm}({C_{p=-1}^2 \oplus A_{1,1}}_{|}.ii)}( gl(1|1)
)$ on proposition 5, one can verify that $R$-matrix
(\ref{Quantiz.29}) satisfies in the following relation:
$$
\sigma \circ \Delta ({\bf X})=R \Delta ({\bf X})
R^{-1}~~~~~~~~~~~~for ~~{\bf X} \in \{ X_1,\cdots, X_4 \}.
$$}
\begin{equation}
R\;=\;\left( \begin{tabular}{cccc}
              $1$ &$0$ &$0$ &$0$ \\
              $0$ &$e^{h}$ &$0$ &$0$ \\
              $0$ &$0$ &$e^{-h}$ &$0$ \\
              $0$ &$0$ &$0$ &$1$ \\
              \end{tabular}
              \right).
 \label{Quantiz.29}
\end{equation}
For the supergroup $GL_h(1|1)$ the $h$-matrix elements $T\in
GL_h(1|1)$ have the form
\begin{equation}
T\;=\;\left( \begin{tabular}{cc}
              $a$ &$\alpha$  \\
              $\beta$ &$b$ \\
              \end{tabular}
              \right),
 \label{Quantiz.30}
\end{equation}
where $a, b$ and $\alpha, \beta$ are bosonic and fermionic
generators, respectively. Now, using  the relation $RT_1T_2=T_2
T_1 R$   in which {\small
\begin{equation} T_1=T\otimes 1=\left(
\begin{tabular}{cccc}
              $a$ &$0$ &$\alpha$ &$0$ \\
              $0$ &$a$ &$0$ &$-\alpha$ \\
              $\beta$ &$0$ &$b$ &$0$ \\
              $0$ &$-\beta$ &$0$ &$b$ \\
              \end{tabular}
              \right),~~~ T_2=1\otimes T_=\left( \begin{tabular}{cccc}
              $a$ &$\alpha$ &$0$ &$0$ \\
              $\beta$ &$b$ &$0$ &$0$ \\
              $0$ &$0$ &$a$ &$\alpha$ \\
              $0$ &$0$ &$\beta$ &$b$ \\
              \end{tabular}
              \right),
 \label{Quantiz.31}
\end{equation}}
one can obtain the commutation relations of the $h$-matrix
elements $T$ as follows:
\begin{equation}\label{Quantiz.32}
\begin{array}{ccl}
a\alpha=\alpha a e^{h}, & \alpha b=e^{-h}b \alpha, & \alpha \beta =-\beta \alpha e^{-2h}, \\
a\beta=e^{-h}\beta a, & \beta b=e^{h}b \beta, & ab =ba,
\end{array} \,\,\,
{\alpha}^2={\beta}^2=0.
\end{equation}
For calculating the superantipode of $a, b, \alpha$ and $\beta$,
we need the quantum superdeterminant of $T$; for this purpose we
use the Gauss decomposition generators \cite{kulish} to obtain
\begin{equation} \label{Quantiz.33}
s\det \nolimits_hT= {a^2}({ab-e^{-h} \beta \alpha})^{-1}.
\end{equation}
Hence, the supercoproduct, supercounit and superantipode of $a, b,
\alpha$ and $\beta$ are determined to be
$$
\Delta(a)\;=a \otimes a + \alpha \otimes
\beta,~~~~~~~\Delta(\alpha)\;=a \otimes \alpha + \alpha \otimes b,
$$
\begin{equation}
\Delta(b)\;=\beta \otimes \alpha + b \otimes
b,~~~~~~~\Delta(\beta)\;=\beta \otimes a + b \otimes \beta,
 \label{Quantiz.34}
\end{equation}
\begin{equation}
\epsilon(a)\;=\;\epsilon(b)\;=\;1,~~~~~~~~~~~~~~~~~\epsilon(\alpha)\;=\;\epsilon(\beta)\;=\;0,
 \label{Quantiz.35}
\end{equation}
$$
S(a)\;=\;({ab-e^{-h} \beta \alpha}){a^{-2}
b^{-1}},~~~~~~~~S(b)\;=\;({ab+e^{-h} \beta \alpha}){a^{-1}
b^{-2}},~~~
$$
\begin{equation}
S(\alpha)\;=\;-e^{-h}\alpha {a^{-1}
b^{-1}},~~~~~~~~~~~~~S(\beta)\;=\;-{ e^{-h}}{a^{-1} b^{-1}} \beta.
 \label{Quantiz.36}
\end{equation}
Now at the end of this section, one can compare our results with
of those of Ref. \cite{Frappat}. It is seen that our results on
propositions 3 and 4 correspond to theorem 0 of \cite{Frappat}
(when $r \rightarrow 1$). Furthermore, our quantum R-matrix
related to the triangular Lie superbialgebra $(gl(1|1) ,
{C_{p=-1}^2 \oplus A_{1,1}}_{|}.ii )$ is different of those of
\cite{Frappat}. Meanwhile, our results on propositions 5 and 6 are
new and different of those of  \cite{Frappat}.

\section{Conclusion}

In this paper, we first classified decomposable Lie superalgebras
of the type $(2 ,2)$. Then, we  obtained all  $gl(1|1)$ Lie
superbialgebras. In this respect, we determined their types
(triangular, quasi-triangular or factorizable) and also classified
Drinfeld superdoubles generated by  the $gl(1|1)$ Lie superalgebra
as a theorem. Using this classification one can investigate super
Poisson-Lie symmetry of the sigma models (specially in the WZW
model) on the $\bf GL(1|1)$ Lie supergroup. Furthermore, one can
construct string cosmology models with super Poisson-Lie symmetry
\cite{cosmology} on the $\bf GL(1|1)$ Lie supergroup. By applying
the procedure of constructing integrable systems \cite{Gould}, we
found a new integrable system on the supersymplectic homogeneous
superspace $OSp(1|2)/U(1)$. Here, we have used  the coboundary
Lie superbialgebra of the type triangular, i.e., $(gl(1|1) ,
{C^2_{p=-1}  \oplus A_{1,1}}_{|}.ii)$. In the same way, one can
construct other integrable systems on the $OSp(1|2)/U(1)$
superspace using the classical $r$-matrices concerning the $(3|2)$
and $(4|4)$-dimensional coboundary  Lie superbialgebras
\cite{{J.z},{J}}. Finally, we quantized $gl(1|1)$ Lie
superalgebra  by using the Lyakhovsky and Mudrov formalism and
for one case obtained quantum R-matrix. Similarly, one can obtain
quantum R-matrices for other Lie superbialgebras. Some of these
remarks are under investigation.

\bigskip
{\it Acknowledgments}:~ This research was supported by a research
fund No. 401.231 from Azarbaijan Shahid Madani university. The
authors would like to thank F. Darabi for carefully reading the
manuscript and useful comments.

\newpage

{\bf \large Appendix A} :~{\bf Isomorphism matrices between the
Manin supertriples based on the $gl(1|1)$}

\vspace{8mm}

\begin{itemize}

\item {\bf $\cal{D}$$sd^{1}_{(4,4)}:$}~~~~~~~~~~~~~~~~{\footnotesize $\Big(gl{(1 | 1)}, I_{(2 , 2)}\Big) \longrightarrow
\Big(gl{(1 | 1)}, C^2_{p=-1}\oplus A_{1,1_{|}}.ii\Big)$}

$$
C=\left( \begin{tabular}{cccccccc}
              $ 1$& $m$& $n$& $ 0$ & $0$& $0$& $ 0$ & $0$ \\
              $0$  & $bc$& $cd-be$& $ 0$ & $0$& $0$& $ 0$ & $0$ \\
              $0$  & $bc$& $abc-be+cd$& $ 0$ & $0$& $0$& $ 0$ & $0$ \\
              $-1$ & $r$& $s$& $ a$ & $0$& $0$& $ 0$ & $0$ \\
              $ 0$ & $0$& $0$& $ 0$ & $b$& $0$& $ 0$ & $d$ \\
               $0$ & $0$& $0$& $ 0$ & $0$& $c$& $e$ & $0$ \\
              $ 0$ & $0$& $0$& $ 0$ & $0$& $0$& $ ac$ & $0$ \\
              $ 0$ & $0$& $0$& $ 0$ & $0$& $0$& $0$ & $ab$ \\
              \end{tabular}
              \right),
$$
$~~~~~~~~~~~~~~a, b, c\in\Re-\{0\};\;     e, d, m, n, r, s\in \Re,
$ \vspace{6mm}

\item {\bf $\cal{D}$$sd^{2\;p}_{(4,4)}:$}\\

{\footnotesize $\Big(gl{(1 | 1)}, B \oplus A \oplus
A_{1,1_{|}}.i\Big) \longrightarrow \Big(gl{(1 | 1)}, B \oplus A
\oplus A_{1,1_{|}}.ii\Big)$}

$$
C=\left( \begin{tabular}{cccccccc}
              $ 1$& $m$& $n$& $ 2$ & $0$& $0$& $ 0$ & $0$ \\
              $0$  & $ad$& $-ad+bc$& $ 0$ & $0$& $0$& $ 0$ & $0$ \\
              $0$  & $0$& $-bc$& $ 0$ & $0$& $0$& $ 0$ & $0$ \\
              $0$ & $r$& $s$& $1$ & $0$& $0$& $ 0$ & $0$ \\
              $ 0$ & $0$& $0$& $ 0$ & $0$& $0$& $ a$ & $b$ \\
               $0$ & $0$& $0$& $ 0$ & $d$& $c$& $-c$ & $-d$ \\
              $ 0$ & $0$& $0$& $0$ & $0$& $c$& $-c$ & $0$ \\
              $ 0$ & $0$& $0$& $ 0$ & $0$& $0$& $a$ & $0$ \\
              \end{tabular}
              \right),
$$
$\;\;\;\;\;a, b, c, d\in\Re-\{0\};\;\;m, n, r, s\in \Re,$
\newpage

{\footnotesize $\Big(gl{(1 | 1)}, B \oplus A \oplus
A_{1,1_{|}}.i\Big) \longrightarrow \Big(gl{(1| 1)},
C^2_{p=1}\oplus A_{1,1_{|}}.i\Big)$}

$$
C=\left( \begin{tabular}{cccccccc}
              $ 1$& $m$& $n$& $ 0$ & $0$& $0$& $ 0$ & $0$ \\
              $0$  & $a(b+d)$& $bc-ad$& $ 0$ & $0$& $0$& $ 0$ & $0$ \\
              $0$  & $a(b+d)$& $-2ab-bc-ad$& $ 0$ & $0$& $0$& $ 0$ & $0$ \\
              $-1$ & $r$& $s$& $ -2$ & $0$& $0$& $ 0$ & $0$ \\
              $ 0$ & $0$& $0$& $ 0$ & $a$& $0$& $ 0$ & $c$ \\
               $0$ & $0$& $0$& $ 0$ & $0$& $b$& $d$ & $0$ \\
              $ 0$ & $0$& $0$& $0$ & $0$& $2b$& $ -2b$ & $0$ \\
              $ 0$ & $0$& $0$& $ 0$ & $2a$& $0$& $0$ & $-2a$ \\
              \end{tabular}
              \right),
$$
$$
\left\{\begin{array}{ll} a+c\neq 0, &\\ b+d\neq 0,&\end{array}
\right. \left\{\begin{array}{ll} a, b\in\Re-\{0\}, &\\ m, n, r, s,
c\in \Re,&\end{array} \right.~~~~~~~~~~~~
$$

\vspace{3mm}

{\footnotesize $\Big(gl{(1 | 1)}, B \oplus A \oplus
A_{1,1_{|}}.i\Big) \longrightarrow \Big(gl{(1| 1)}, C^2_{p}\oplus
A_{1,1_{|}}.i\Big)$}

$$
C=\left( \begin{tabular}{cccccccc}
              $ 1$& $m$& $n$& $ 2$ & $0$& $0$& $ 0$ & $0$ \\
              $0$  & $ad$& $-ad+bc$& $ 0$ & $0$& $0$& $ 0$ & $0$ \\
              $0$  & $ad$& $-pbc-ad$& $ 0$ & $0$& $0$& $ 0$ & $0$ \\
              $-1$ & $r$& $s$& $ p-1$ & $0$& $0$& $ 0$ & $0$ \\
              $ 0$ & $0$& $0$& $ 0$ & $0$& $0$& $ d$ & $c$ \\
               $0$ & $0$& $0$& $ 0$ & $a$& $b$& $-b$ & $-a$ \\
              $ 0$ & $0$& $0$& $0$ & $0$& $(1+p)b$& $ -(1+p)b$ & $0$ \\
              $ 0$ & $0$& $0$& $ 0$ & $0$& $0$& $(1+p)d$ & $0$ \\
              \end{tabular}
              \right),
$$
$\;\;\;\;\;a, b, c, d\in\Re-\{0\};\;\;m, n, r, s\in \Re,$
\vspace{2mm}

{\footnotesize $\Big(gl{(1 | 1)}, B \oplus A \oplus
A_{1,1_{|}}.i\Big) \longrightarrow \Big(gl{(1| 1)},
C^2_{\frac{1}{p}}\oplus A_{1,1_{|}}.ii\Big)$}

$$
C=\left( \begin{tabular}{cccccccc}
              $ 1$& $m$& $n$& $ 2$ & $0$& $0$& $ 0$ & $0$ \\
              $0$  & $ac$& $-ac-bd$& $ 0$ & $0$& $0$& $ 0$ & $0$ \\
              $0$  & $ac$& $\frac{1}{p}bd-ac$& $ 0$ & $0$& $0$& $ 0$ & $0$ \\
              $-1$ & $r$& $s$& $ \frac{1}{p}-1$ & $0$& $0$& $ 0$ & $0$ \\
              $ 0$ & $0$& $0$& $ 0$ & $0$& $0$& $ a$ & $b$ \\
               $0$ & $0$& $0$& $ 0$ & $c$& $-d$& $d$ & $-c$ \\
              $ 0$ & $0$& $0$& $0$ & $0$& $\frac{-(1+p)d}{p}$& $ \frac{(1+p)d}{p}$ & $0$ \\
              $ 0$ & $0$& $0$& $ 0$ & $0$& $0$& $\frac{(1+p)a}{p}$ & $0$ \\
              \end{tabular}
              \right),
$$
$~~~~a, b, c, d\in\Re-\{0\};\;\;m, n, r, s\in \Re,$

 \vspace{4mm}

\item {\bf $\cal{D}$$sd^{3\;\epsilon_1, \epsilon_2}_{(4,4)}:$}\\

{\footnotesize $\Big(gl{(1 | 1)}, (B\oplus (A_{1,1}+
A))_{\epsilon_1 =1}.i\Big) \longrightarrow \Big(gl{(1 | 1)},
(B\oplus (A_{1,1}+ A))_{\epsilon_2 =-1}.i\Big)$}

$$
C=\left( \begin{tabular}{cccccccc}
              $ 1$& $m$& $n$& $ 0$ & $0$& $0$& $ 0$ & $0$ \\
              $0$  & $a^2$& $0$& $ 0$ & $0$& $0$& $ 0$ & $0$ \\
              $0$  & $0$& $-a^2$& $ 0$ & $0$& $0$& $ 0$ & $0$ \\
              $0$ & $b$& $c$& $ -1$ & $0$& $0$& $ 0$ & $0$ \\
              $ 0$ & $0$& $0$& $ 0$ & $a$& $0$& $ 0$ & $0$ \\
               $0$ & $0$& $0$& $ 0$ & $0$& $a$& $0$ & $0$ \\
              $ 0$ & $0$& $0$& $0$ & $0$& $0$& $ -a$ & $0$ \\
              $ 0$ & $0$& $0$& $ 0$ & $0$& $0$& $0$ & $-a$ \\
              \end{tabular}
              \right),
$$
$\;\;\;\;\;a \in\Re-\{0\};\;\;m, n, b, c\in \Re,$
\vspace{12mm}

{\footnotesize $\Big(gl{(1 | 1)}, (B\oplus (A_{1,1}+
A))_{\epsilon_1}.i\Big) \longrightarrow \Big(gl{(1 | 1)}, (B\oplus
(A_{1,1}+ A))_{\epsilon_2}.ii\Big)$}

$$
C=\left( \begin{tabular}{cccccccc}
              $ -1$& $m$& $n$& $ 0$ & $0$& $0$& $ 0$ & $0$ \\
              $0$  & $-a^2$& $0$& $ 0$ & $0$& $0$& $ 0$ & $0$ \\
              $0$  & $0$& $\frac{\epsilon_2}{\epsilon_1}a^2$& $ 0$ & $0$& $0$& $ 0$ & $0$ \\
              $0$ & $b$& $c$& $ \frac{\epsilon_2}{\epsilon_1}$ & $0$& $0$& $ 0$ & $0$ \\
              $ 0$ & $0$& $0$& $ 0$ & $0$& $-a$& $ 0$ & $0$ \\
               $0$ & $0$& $0$& $ 0$ & $a$& $0$& $0$ & $0$ \\
              $ 0$ & $0$& $0$& $0$ & $0$& $0$& $ 0$ & $\frac{\epsilon_2}{\epsilon_1}a$ \\
              $ 0$ & $0$& $0$& $ 0$ & $0$& $0$& $-\frac{\epsilon_2}{\epsilon_1}a$ & $0$ \\
              \end{tabular}
              \right),
$$
$\;\;\;\;\;a \in\Re-\{0\};\;\;m, n, b, c\in \Re,$

\newpage

\item {\bf $\cal{D}$$sd^{4\;\epsilon_1, \epsilon_2}_{(4,4)}:$}\\

{\footnotesize $\Big(gl{(1 | 1)}, (C^3+ A)_{\epsilon_1=1}.i\Big)
\longrightarrow \Big(gl{(1 | 1)}, (C^3+
A)_{\epsilon_2=-1}.i\Big)$}

$$
C=\left( \begin{tabular}{cccccccc}
              $1$& $m$& $n$& $ 0$ & $0$& $0$& $ 0$ & $0$ \\
              $0$  & $-ab^2$& $0$& $ 0$ & $0$& $0$& $ 0$ & $0$ \\
              $0$  & $0$& $-a^2b^2$& $ 0$ & $0$& $0$& $ 0$ & $0$ \\
              $0$ & $c$& $d$& $a$ & $0$& $0$& $ 0$ & $0$ \\
              $ 0$ & $0$& $0$& $ 0$ & $b$& $0$& $ 0$ & $0$ \\
               $0$ & $0$& $0$& $ 0$ & $0$& $-ab$& $0$ & $0$ \\
              $ 0$ & $0$& $0$& $0$ & $0$& $0$& $ -a^2b$ & $0$ \\
              $ 0$ & $0$& $0$& $ 0$ & $0$& $0$& $0$ & $ab$ \\
              \end{tabular}
              \right),
$$
$\;\;\;\;\;a, b \in\Re-\{0\};\;\;m, n, c, d\in \Re,$

\vspace{4mm}

{\footnotesize $\Big(gl{(1 | 1)}, (C^3+ A)_{\epsilon_1}.i\Big)
\longrightarrow \Big(gl{(1 | 1)}, (C^3+ A)_{\epsilon_2}.ii\Big)$}

$$
C=\left( \begin{tabular}{cccccccc}
              $ -1$& $c$& $d$& $ 0$      & $0$& $0$& $ 0$ & $0$ \\
              $0$  & $ab$& $0$& $ 0$ & $0$& $0$& $ 0$ & $0$ \\
              $0$  & $0$& $-\frac{\epsilon_2 a^2}{\epsilon_1}$& $ 0$ & $0$& $0$& $ 0$ & $0$ \\
              $ 0$ & $e$& $f$& $ \frac{\epsilon_2 a}{\epsilon_1 b}$ & $0$& $0$& $ 0$ & $0$ \\
              $ 0$ & $0$& $0$& $ 0$ & $0$& $a$& $ 0$ & $0$ \\
               $0$ & $0$& $0$& $ 0$ & $b$& $0$& $ 0$ & $0$ \\
              $ 0$ & $0$& $0$& $0$ & $0$& $0$& $ 0$ & $\frac{\epsilon_2 a}{\epsilon_1}$ \\
              $ 0$ & $0$& $0$& $ 0$ & $0$& $0$& $ \frac{\epsilon_2 a^2}{\epsilon_1 b}$ & $0$ \\
              \end{tabular}
              \right),
$$
$\;\;\;\;\;a, b\in\Re-\{0\};\;\;c, d, e, f\in
              \Re.$
\vspace{4mm}

\end{itemize}


\end{document}